\documentclass[useAMS, usenatbib, a4paper]{mn2e}

\usepackage{mathrsfs}
\usepackage[space]{grffile}
\usepackage[T1]{fontenc}
\usepackage{aecompl}
\usepackage[dvipsnames]{xcolor}
\usepackage{times}
\usepackage{subfigure}
\usepackage{amsmath, amssymb}
\usepackage{amsfonts}
\usepackage{graphicx}
\usepackage{multirow}
\usepackage{hyperref}
\usepackage{url}
\usepackage{bm}
\usepackage{ulem} 
\usepackage[abs]{overpic}
\usepackage{tikz}
\usepackage{todonotes}
\usepackage{times}
\usepackage{amssymb}
\usepackage{amsmath}
\usepackage{graphicx,color}
\usepackage{bmpsize}
\usepackage{epstopdf}
\usepackage{adjustbox}
\usepackage{multicol}
\usepackage{rotate}

\newcommand{\vect}[1]{\mbox{\boldmath ${#1}$}}
\newcommand {\apgt} {\ {\raise-.5ex\hbox{$\buildrel>\over\sim$}}\ }
\newcommand {\aplt} {\ {\raise-.5ex\hbox{$\buildrel<\over\sim$}}\ }

\def\myputfigure#1#2#3#4#5%
{\vskip#5pt\makebox[0pt]{\hskip#2in
\includegraphics[width=#3\textwidth]{#1}}\vskip#4pt\hfill}

\newcommand{\cmfast}{\textsc{\small 21CMFASTv3}}
\newcommand{\pyobs}{\textsc{\small pyObs21}}
\newcommand{\cmmc}{\textsc{\small 21cmmc}}
\newcommand{\cmsense}{\textsc{\small 21CMsense}}

\newcommand{\bifft}{\textsc{\small BiFFT}}

\newcommand{\hi}{\mathrm{H}\textsc{i} }

\newcommand{\Tvir}{T_{\mathrm{vir}}}
\newcommand{\Rmax}{ R_{\mathrm{max}} }

\newcommand{\cmpc}{\mathrm{ cMpc}}
\newcommand{\invcmpc}{\mathrm{ cMpc}^{-1}}

\title[EoR parameter estimation with the 21-cm bispectrum.]
{Epoch of reionization parameter estimation with the 21-cm bispectrum}
\author[C. A. Watkinson, B. Greig, A. Mesinger]
{Catherine ~A.~Watkinson$^{1}$\thanks{Email: \href{mailto:catherine.watkinson@gmail.com}
{\protect\nolinkurl{catherine.watkinson@gmail.com}}}, Bradley Greig$^{2,3}$, Andrei Mesinger$^4$\\
$^1$School of Physics and Astronomy, Queen Mary University of London, Mile End Road, London E1 4NS, UK \\
$^2$School of Physics, University of Melbourne, Parkville, VIC 3010, Australia\\
$^3$ARC Centre of Excellence for All-Sky Astrophysics in 3 Dimensions (ASTRO 3D), University of Melbourne, VIC 3010, Australia \\
$^4$Scuola Normale Superiore, Piazza dei Cavalieri 7, I-56126 Pisa, Italy \\}
\date{\today}

\pubyear{\number\year}

\begin{document}
\maketitle

\begin{abstract}
We  present  the  first  application  of  the  isosceles  bispectrum  to MCMC parameter inference from the cosmic 21-cm signal.  We extend the MCMC sampler \cmmc\ to use the fast bispectrum code, \bifft, when computing the likelihood.
We create mock 1000h observations with SKA1-low, using \pyobs\,to account for uv-sampling and thermal noise.
Assuming the spin temperature is much higher than that of the CMB, we consider two different reionization histories for our mock observations: fiducial and late-reionization. 
For both models we find that bias on the inferred parameter means and 1-$\sigma$ credible intervals can be substantially reduced by using the isosceles bispectrum (calculated for a wide range of scales and triangle shapes) together with the power spectrum (as opposed to just using one of the statistics). We find that making the simplifying assumption of a Gaussian likelihood with a diagonal covariance matrix does not notably bias parameter constraints for the three-parameter reionization model and basic instrumental effects considered here. This is true even if we use extreme (unlikely) initial conditions which would be expected to amplify biases.
We also find that using the cosmic variance error calculated with Monte-Carlo simulations using the fiducial model parameters whilst assuming the late-reionization model for the simulated data also does not strongly bias the inference.  This implies we may be able to sparsely sample and interpolate the cosmic variance error over the parameter space, substantially reducing computational costs.  All codes used in this work are publicly-available.
\end{abstract}

\section{Introduction}\label{sec:intro}

The Square Kilometre Array\footnote{\sloppy The Square Kilometre Array
\url{http://www.skatelescope.org/} and \url{https://astronomers.skatelescope.org/wp-content/uploads/2016/05/SKA-TEL-SKO-0000002_03_SKA1SystemBaselineDesignV2.pdf}} aims to detect the high-redshift 21-cm line of neutral hydrogen. It is projected to produce high precision maps at a wide range of redshifts.
These maps can be used to infer the properties of early generations of stars and galaxies as they influence the intergalactic medium (IGM) via coupling, heating and ionizations \citep{Dewdney2016}.
The phase change in the Universe's ionization state induced by the latter process is called the Epoch of Reionization (EoR).

Numerous studies have predicted great benefits from using
higher-order statistics such as the bispectrum in our analysis of such datasets.
For example, \citealt{Shimabukuro2016a, Majumdar2017, Watkinson2018, Hutter2019} and \citealt{Gorce2019} show that,
due to the non-Gaussian nature of the signal, additional information is contained in higher order statistics, which unlike the power spectrum are sensitive to non-Gaussian structure in a dataset.
In particular, \citealt{Shimabukuro2016a} perform a Fisher forecast and find that using the equilateral bispectrum in addition to the power spectrum substantially shrinks the credible limits of the parameters of a three-parameter EoR model compared to those resulting from using the power spectrum alone.

Furthermore, it appears that the error due to instrumental noise is not as large
as one might naively expect;
see for example, \citealt{Yoshiura2015, Watkinson2018}, and \citealt{Trott2019}. This is because Gaussian distributed noise has
zero bispectrum so that it is only the statistical fluctuations of the noise bispectrum that
contributes to the measured error on the bispectrum \citep{Yoshiura2015}.

The Fisher analysis of \citealt{Shimabukuro2016a}, whilst an important first step towards understanding the improvements gained in performing parameter estimation with the bispectrum, likely underestimates the credible limits associated with each parameter.
This is because a parameter's covariance matrix is only accurately described by the inverse of the Fisher matrix if the errors on the measured quantities are perfectly Gaussian (i.e. the likelihood surface is Gaussian around the maximum likelihood point), which is not a given for even the 21-cm power spectrum. It has also been shown that the covariance predicted by a Fisher forecast, by the Cramer-Rao bound, provides the smallest possible attainable error, i.e. it provides a lower limit \citep{Fisher1935, Cramer1946, Rao1945, Tegmark1996}.
In this paper we take the work of \citealt{Shimabukuro2016a} a step further by adding the isosceles bispectrum (in which we include a wide range of triangle configurations in addition to the equilateral) within a Monte-Carlo Markov Chain (MCMC) parameter estimation framework, building on the established 21-cm MCMC code \cmmc\,\citep{Greig2015a, Greig2017, Park2018}.

Section~\ref{sec:like_bg} describes our bispectrum likelihood and the methods used to simulate instrumental effects and measure the bispectrum.
In Section~\ref{sec:ideal} we look at an idealised case with no instrumental effects or sample variance to see the maximal achievable improvement to the parameter constraints when combining the bispectrum and power spectrum.
In Section~\ref{sec:sv} we compare analytic approximations to the sample-variance error with the true sample-variance error calculated using Monte-Carlo (MC) methods.
We will show in this section that assuming a sample-variance error that is a fixed percentage of the statistics in any given bin is a very poor approximation, as is propagating the power-spectrum sample-variance error onto the bispectrum assuming Gaussianity.
In Section~\ref{sec:mainresults} we present our main analysis that include instrumental effects ($uv$ sampling \& noise) and sample-variance. We will show in this section that using the bispectrum in combination with the power spectrum reduces the bias (and in some cases the credible intervals) on all parameters relative to that of the power-spectrum only analysis. This is true regardless of how likely is the realization of the "true" Universe (i.e. if the initial conditions are outliers) or its reionization history.

\section{Inclusion of instrumental effects and bispectrum likelihood to \cmmc}\label{sec:like_bg}

For the purposes of this analysis, we modify the latest version of \cmmc: an MC sampler of \cmfast\,(a python-wrapped, semi-numerical simulation of the 21-cm signal at high redshifts) \citep{Murray2020}.
\cmmc\,can be downloaded from \url{https://github.com/21cmFAST/21CMMC}, and is detailed in: \citealt{Greig2015a} (which describes the first implementation that used a three-parameter model for reionization), \citealt{Greig2017} (which extends sampling to parameters responsible for heating and Lyman-$\alpha$ coupling effects), and \citealt{Park2018} (which introduces mass dependence to the star formation rates and escape fraction of ionizing radiation, as well as luminosity functions).  The latest version of \cmmc\ has the option of using either the EMCEE or Multinest samplers; here we use EMCEE which is an  Affine-invariant, openMP-parallelized MCMC sampler (for more details see \url{https://emcee.readthedocs.io/en/stable/}) \citep{Goodman2010, Foreman-Mackey2013}.

\cmfast\,is a standalone code for computing 3D realizations of the 21-cm signal and its component fields.  Sampling Gaussian initial conditions, it uses Lagrangian perturbation theory to generate density and velocity fields (e.g. \citealt{Bernardeau2001}); then using a combination of excursion set \citep{Furlanetto2004a} and lightcone integration it generates ionization and temperature fields.
We refer the interested reader to \citealt{Mesinger2007} and \citealt{Signal2010} for details, as well as to the extensive documentation associated with the code itself available at \url{https://github.com/21cmfast/21cmFAST}.

For this demonstrative work, we use the simplest, three parameter reionization model (as described in \citealt{Greig2015a}), and assume the spin temperature exceeds the CMB temperature.  We also compute our summary statistics from coeval cubes, instead of lightcones.\footnote{A coeval cube is a datacube that has been simulated using a fixed cosmological time throughout. A lightcone dataset is one in which the simulated epoch evolves with frequency (or redshift), i.e. each slice along the $z$-axis represents a different cosmological time. 
} These choices keep the analysis time to a minimum facilitating the ability to experiment with different aspects of the analysis whilst still being informative.  In future work, we will relax these assumptions.

The parameters that we vary in our analysis are: 

\begin{itemize}
\item $\zeta = 
f_{\mathrm{esc}}
\,f_*\,N_{\gamma/b}
\,
(1+n_{\mathrm{rec}})^{-1}$ which is the ionizing efficiency of galaxies.  Here $f_{\mathrm{esc}}$ is the escape fraction of ionizing photons, $N_{\gamma/b}$ is the number of ionizing photons produced per baryon in stars, and  $n_{\mathrm{rec}}$ is the cumulative number of IGM recombinations per baryon. This is assumed to be a constant, and a region is deemed to be ionized if the collapsed fraction within that region is greater than or equal to $\zeta^{-1}$.
Increasing $\zeta$ therefore speeds up the EoR.
\item $\Tvir$ is the minimum virial temperature needed for halos to host star-forming galaxies (determined by cooling and feedback mechanisms that allow star formation). Smaller $\Tvir$ means star formation is possible in lower-mass halos that are less biased. Thus reducing $\Tvir$ results in an earlier EoR, characterized by smaller, more uniformly-distributed cosmic HII regions.
\item $\Rmax$ defines the maximum distance a photon can travel in an ionized IGM before it encounters a recombined atom.  This effective parameter can loosely be related to a characteristic mean free path (c.f. \citealt{Furlanetto2005} and \citealt{Sobacchi2014}).
As $\Rmax$ is only relevant when it is smaller than the typical HII region size, reducing it extends the late stages of the EoR without impacting the early stages.
\end{itemize}

We make the assumption that the power spectrum and bispectrum measurements are independent (from each other and between each $k$ bin for the power spectrum or triangle configuration for the bispectrum). We also assume independence of these statistics at each redshift. This allows us to approximate the total likelihood using a simple sum over $\chi^2$ values.  Specifically, we take $\mathrm{ln}\,\mathcal{L}(\theta|d) = - \sum_{ij} (d_{ij} - m_{ij})/(2\,\sigma_{ij})$ where the indices denote redshift and statistical bins, i.e. each $ij$ corresponds to a the measurement of a single power spectrum or bispectrum bin (from the data $d_{ij}$ or model $m_{ij}$) at one of the redshift bins under consideration. For the main results of this paper we pre-compute $\sigma_{ij}$ by forward simulating the fiducial model, each time varying the initial seed of the simulation to account for sample variance error, and including a random realisation of instrumental noise.
The standard deviation we use in this study is calculated using 2000 such Monte-Carlo (MC) samples of the power spectrum and bispectrum in each bin (although it is worth noting that the error estimate has mostly converged by 1000 iterations).

We ignore the contribution to the power spectrum and bispectrum for $k$ modes that fall outside of the range $0.1\le k \le 1.0$ cMpc$^{-1}$.
The lower $k$ cut is motivated by avoiding modes that are likely to suffer from corruption due to foreground leakage, and the upper cut excludes modes that will suffer from the effects of shot noise \citep{Greig2015a}.
For the bispectrum this means that if any one of the three $k$-vectors that form a given triangle configuration fall outside of this range, then the configuration is excluded from our likelihood calculation.

We set our \textbf{fiducial model} parameter values as
$\zeta = 30.0$, $\log\Tvir = 4.7$ and $R_{\mathrm{max}} = 15$.
We also consider a \textbf{late reionization model} with $\zeta = 17.0$, $\log\Tvir = 5$ and $R_{\mathrm{max}} = 10$.
We initialise the core of \cmmc\,to simulate coeval cubes at $z = [6.3, 7, 8, 9]$,
chosen to sample a range of ionized fractions,
with our redshifts corresponding to $x_{\hi} =  [0.13, 0.33, 0.62, 0.79]$
for our fiducial model and $x_{\hi} =  [0.70, 0.80, 0.89, 0.94]$ for our late reionization model.
Note that our late reionization model is not picked as a realistic model, it is selected somewhat arbitrarily to provide a test case that is quite different to the fiducial model.\footnote{The ionized fractions we quote are for our "standard" seed, which we discuss in section \ref{sec:sv}.}
We use the same prior ranges as \citealt{Greig2015a}, i.e. $10\le\zeta\le30$, $4\le \Tvir \le 6$ and $5\le R_{\mathrm{max (bubble)}}\le 20$.
Our coeval cubes are $128^3$ and (256 Mpc)$^3$ in dimension, chosen to keep both sample variance and analysis time to acceptable levels \citep{Iliev2014, Kaur2020}.

\subsection{\bifft\,- a fast code for measuring the bispectrum}

The bispectrum is defined as the Fourier transform of the three-point correlation function (which measures excess probability as a function of three points in real space). It can be written as
\begin{equation}
\begin{split}
(2\pi)^3 B(\vect{k}_1, \vect{k}_2, \vect{k}_3) \delta^{\mathrm{D}}(\vect{k}_1 + \vect{k}_2 +  \vect{k}_3  )=
\langle \Delta(\vect{k}_1)\Delta(\vect{k}_2)\Delta(\vect{k}_3)\rangle \,,\\
\label{eqn:bispec_def}
\end{split}
\end{equation}
where $\delta^{\mathrm{D}}(\vect{k}_1 + \vect{k}_2 + \vect{k}_3)$ is the Dirac-delta function.
Accordingingly, the bispectrum is a function of three $k$ vectors that form a closed triangle, often referred to (as we will from here on) as a triangle configuration.
It is necessary to perform some kind of averaging when measuring the bispectrum to beat down statistical noise. As is common in bispectrum and power spectrum analysis, we choose to perform spherical averaging, i.e. our bispectrum measurements are functions of triangle shape and size only, not orientation.

The bispectrum is the lowest order polyspectra that is sensitive to non-Gaussian information, or structure, in a dataset.
For a nice description of what real-space structures different $k$-space triangle configurations are sensitive to see \citealt{Lewis2011, Watkinson2018} and \citealt{Hutter2019} (see in particular Figure 1).

Due to computational limitations, the bispectrum is often overlooked in forward-modeling frameworks.  Naively, it requires multiple nested loops to find the $k$-space pixels that form closed triangles of the desired shape and size. However, there are methods that make the calculation tractable for many applications. One of these is to use Fast-Fourier Transforms to enforce the Dirac-delta function in equation \ref{eqn:bispec_def} \citep{Scoccimarro2015, Sefusatti2015}. 
\bifft\,is
a python package that wraps a C implementation of the Fourier-transform bispectrum method,
described in \citealt{Watkinson2017} and publicly available from
\url{https://bitbucket.org/caw11/bifft}.
It's very fast, taking only a few seconds per triangle configuration on a MacBookPro (2.3GHz i9 intel core, 16Gb RAM) for a datacube of size $256^3$.
This method is extensively described in \citealt{Watkinson2017} and \citealt{Watkinson2020}.

Throughout we will normalise out the amplitude of the bispectrum to isolate the non-Gaussian information:
\begin{equation}
b(k_1, k_2, k_3) = \frac{B(k_1, k_2, k_3)}{\sqrt{(k_1\,k_2\,k_3)^{-1}\,P(k_1)\,P(k_2)\,P(k_3)}}\,.\\
\label{eqn:normB}
\end{equation}
Equation~\ref{eqn:normB} is commonly applied in signal processing, see for example \citealt{Hinich1968, Kim1978, Hinich1995} and \citealt{Hinich2005}. It has also been argued by \citealt{BrillingerD.R.Rosenblatt1967} that Equation~\ref{eqn:normB} is the optimal normalisation for the bispectrum.
The findings of \citealt{Watkinson2018} support this claim for 21-cm datasets.
Note also that the normalised bispectrum is not a direct function of the power spectrum and so linearly combining its likelihood contribution with that of the power spectrum is not an unreasonable choice.

\subsection{$uv$ sampling and noise generation with \pyobs}\label{sec:pyobs21}
In order to carry out our investigation we wrote \pyobs\,(which can be used as a bolt-on module for \cmmc\,or \cmfast)
to apply $uv$ sampling and add Gaussian random noise (with standard deviation
based on \cmsense\,calculations) to a 21-cm brightness-temperature coeval simulation.\footnote{\pyobs\,can be used for lightcone data if it is chunked into cubes, but since \pyobs\,assumes a fixed redshift in translating the $uv$ sampling of the instrument to simulation co-ordinates it is not the ideal tool for use with lightcones.}

The established code \cmsense\,outputs the noise and sample-variance error of the spherically-averaged power spectrum as a function of $k$.
\pyobs\,relies on an adapted version of calc\_sense.py from \cmsense\,which instead outputs a file containing the $k_x, k_y, k_z$ (in cMpc$^{-1}$) corresponding to the instrument's $uv$ sampling and bandwidth associated with the simulation dimensions, along with the noise power spectrum associated with each $uv$ sample.
\cmsense\,is described extensively in \citealt{Pober2013} and \citealt{Pober2014}. We assume optimistically that foregrounds are fully removed and assume a track scan mode of operation.
On the first call to \pyobs, a maskfile of the same dimension as the \cmfast\,simulation is created containing the noise power in each pixel (the noise in pixels that are repeat samples are combined coherently using inverse-covariance weighting) and
zeroed where there are no $uv$-samples.
Once the $uv$-noise maskfile is written to file, \pyobs\,accesses it each time it is called, zeroes any unsampled pixels in the cosmological simulation and adds a random sample of Gaussian noise to each pixel (based on the noise power in the corresponding $uv$-noise maskfile pixel).

By working in simulation co-ordinates (i.e. cMpc) and creating the uv-noise maskfile on the first call,
\pyobs\,is very quick, making it suitable for MC calculations, including
calculating instrumental error on any statistic (that is in itself also relatively quick to compute).
This approach is approximate in that it ignores the evolution of the
$uv$ sampling along the line of sight. It also ignores the effect of the primary beam,
effectively assuming the field size is small enough to not be affected by this (which for the box sizes simulated here is not unreasonable) or that the primary beam as been corrected for.
The SKA noise level produced by this \pyobs\,(using the central region from the current design for the SKA-Low phase 1 telescope model and assuming 1000 hours observation time) is consistent with that predicted by \citealt{Mellema2013} and \citealt{Koopmans2015}. The SKA1-Low
details and antenna locations used for our noise calculations are based on the latest SKA configuration coordinates\footnote{The SKA antenna positions we use are given by the central region antenna positions of \url{https://astronomers.skatelescope.org/wp-content/uploads/2016/09/SKA-TEL-SKO-0000422_02_SKA1_LowConfigurationCoordinates-1.pdf}} (central region) and \citealt{Dewdney2016}.

\section{Parameter recovery using the isosceles bispectrum for an idealised case}\label{sec:ideal}


In this work we only consider the isosceles configuration as a function of angle
between k1 and k2, and for a range of scales.
Our range of isosceles triangles span shapes from squeezed to stretched, and should therefore be able to pick up a large range of non Gaussian structures in the 21-cm maps.
We refer the reader to Section 3 of \citealt{Watkinson2018} and
\citealt{Lewis2011} for discussions of the types of structures that various configurations
are sensitive to, as well as to the results of \citealt{Majumdar2017} for verification
that the isosceles configuration captures  key features of reionization maps.

In this section we compare the parameter constraints achieved when using the isosceles bipectrum (for $k_1 = k_2 = [0.12\invcmpc, 0.3\invcmpc, 0.7\invcmpc, 0.98\invcmpc]$) and for $\theta/\pi = [0.01, 0.05, 0.1, 0.2, 0.33, 0.4, 0.5, 0.6, 0.7, 0.85, 0.95]$ (where $\theta$ is the internal angle to $\vect{k}_1 + \vect{k}_2 $), the power spectrum, and a combination of the two statistics.\footnote{For all the statistics we consider, we disregard contribution from any $k$ modes that fall outside of the range $k_f<k<k_{\mathrm{nyq}}$ where $k_f=2\pi/L$ is the fundamental $k$ scale and $L$ is the length of a side of the simulation, and $k_{\mathrm{nyq}} = 1.0/2.0*N*k_f$ where $N$ is the resolution on a side. A consequence of these restrictions is that not all $\theta$ bins will be included for larger values of $k_1$.}  
To do so we assume a best case scenario of negligible instrumental
effects, perfect foreground removal, and negligible sample variance.
In practice this involves running analysis on the raw coeval cubes produced by \cmfast\,and assuming the same random seed for the data and model.  We also include a modelling uncertainty of 15\% of the modelled statistics, as is default in \cmmc).

\begin{figure}
  \centering
    $\renewcommand{\arraystretch}{-0.75}
    \begin{array}{c}
      \includegraphics[trim=0.25cm -0.5cm -0.6cm 0.0cm, clip=true, scale=0.5]{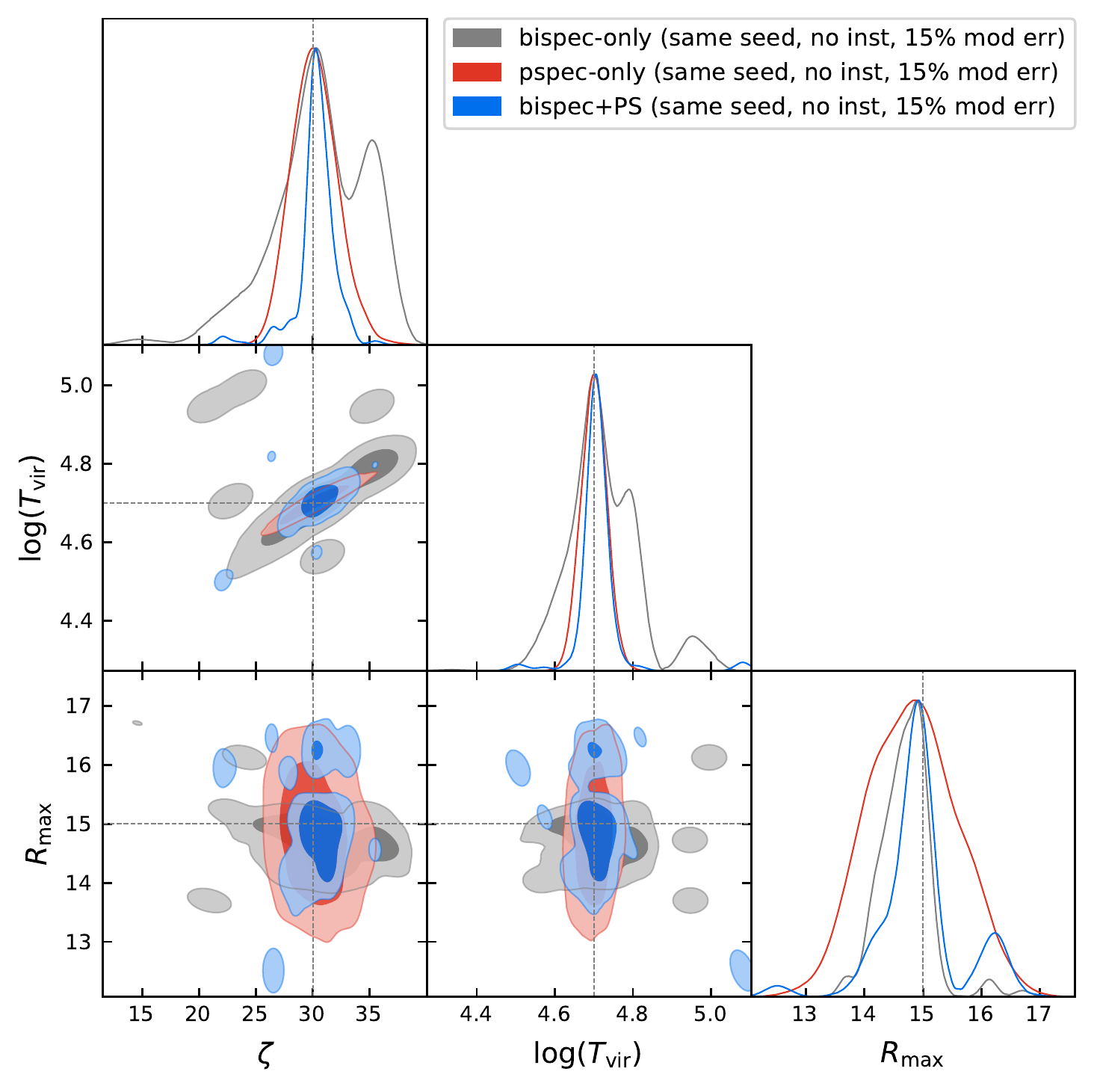}\\
      \includegraphics[trim=1.5cm 0.0cm 0.0cm 3cm, clip=true, scale=0.275]{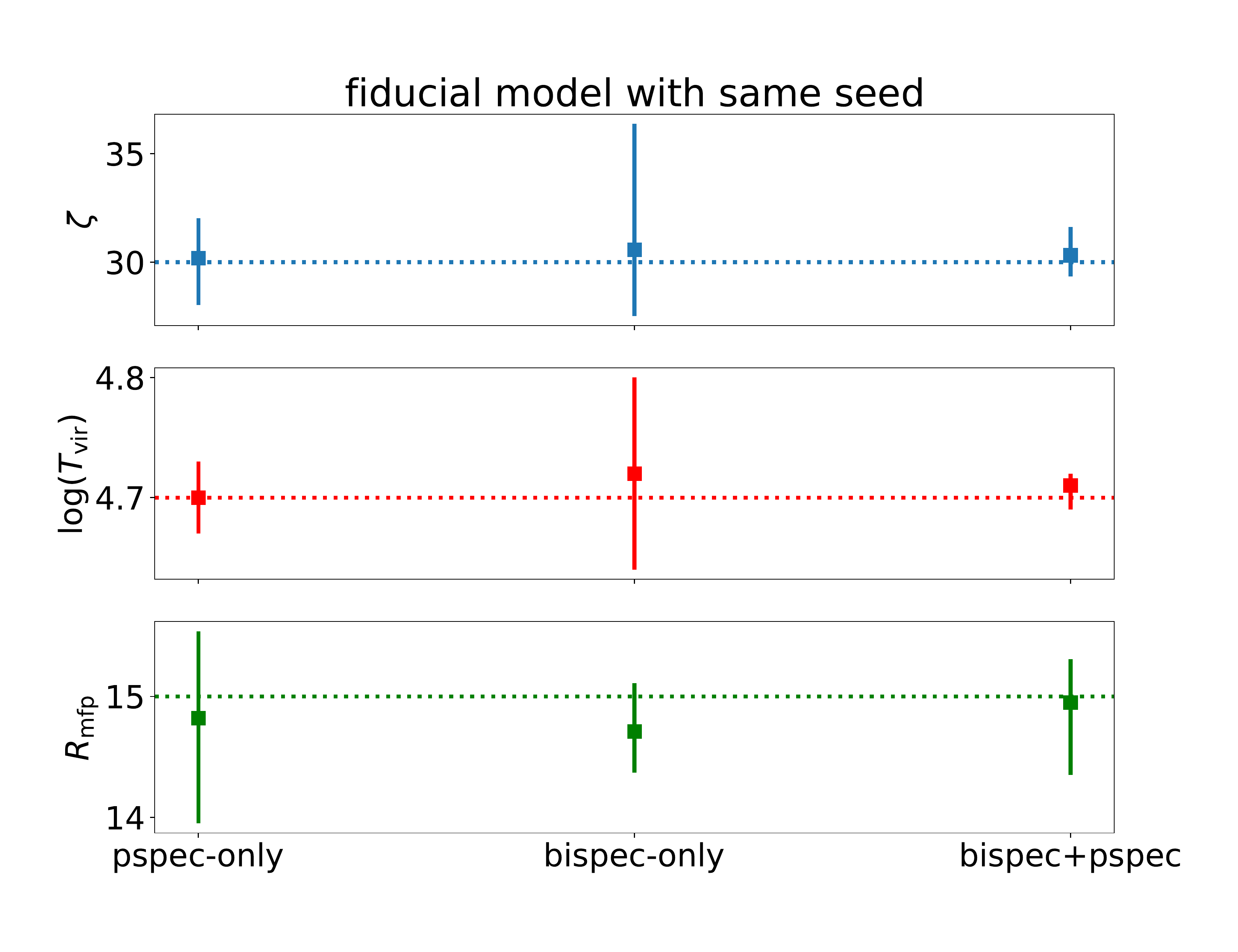}\\
        \end{array}$
  \caption{Corner plot (top)
  for a likelihood based on the spherically-averaged isosceles bispectrum (bispec-only; grey), power spectrum (pspec-only; red),
  and power spectrum + bispectrum (bispec+pspec; blue).
  The bottom plot shows the mean +/- 68\% credible intervals for each parameter.
  All assume a
  best-case scenario of no instrumental effects or foregrounds and use the same random seed for our models and data.
  In this and all the figures that follow,
  our simulations have dimensions of $128^3$ pixels and $(256 \cmpc)^3$ and redshifts simulated
  are $z = [6.3, 7, 8, 9]$.
  For our bispectrum likelihood, we use the isosceles triangle configurations for 11 linearly spaced $\theta$ bins and for $k_1=k_2=[0.12\invcmpc, 0.3\invcmpc, 0.7\invcmpc, 0.98\invcmpc]$ (where $\theta$ is the internal angle to $\boldsymbol{k}_1 + \boldsymbol{k}_2$).
  We see the power spectrum in such a case does a good job of constraining the data but constraints
  are improved by the inclusion of the bispectrum.
 }
  \label{fig:3param_bestcase}
\end{figure}

The corresponding corner plot for the three parameter model is shown in Figure \ref{fig:3param_bestcase}.  Darker/lighter shading encloses 68\%/95\% of the credible limits.  Different colors indicate different statistics used for computing the likelihood: (i) bispectrum is shown with grey; (ii) power spectrum is shown with red; and (iii) bispectrum + power spectrum is shown with blue.

Under these idealised conditions the \textbf{power spectrum only (pspec-only)} statistic
results in tight, unbiased constraints, which can be seen in the bottom of Figure \ref{fig:3param_bestcase} where we plot the marginal statistics, i.e. the marginalised posterior's mean +/- the 68\% upper and lower credible limits.
As in \citealt{Greig2015a}, we see a moderate degeneracy between the ionizing efficiency and the Virial temperature. This is because both parameters effect the timing of reionization; for example, both a high virial temperature and a low ionizing
efficiency will delay and slow the progress of reionization.  The epoch of heating, ignored in this exploratory work, should break this degeneracy (e.g. \citealt{Greig2017}).

We find the pspec-only statistic generally results in tighter constraints than the \textbf{bispectrum only (bispec-only)} statistic.
This tells us that even in the idealized scenario, the amplitude of the signal is more informative than the non-Gaussian information alone (at least for our fiducial model).
However, the credible intervals of $R_\mathrm{max}$ are reduced by a factor of 0.47 relative to the pspec-only case (see also \citealt{Shaw2020}).
This is because $\Rmax$ (by applying a hard limit beyond which photons from a source will cease to be effective at ionizing the IGM) induces structural features to which the bispectrum is particularly sensitive.

When we combine the bispectrum with the power spectrum, the additional information from the non-Gaussianities in the maps greatly reduces the degeneracies of the credible limits for all the parameters.
This corresponds to a shrinkage of the credible intervals by a factor of 0.70, 0.50, 0.60 for $\zeta$, $\mathrm{log}(T_\mathrm{vir})$, and $R_\mathrm{max}$ respectively (with respect to those of pspec-only).
Although we note that the marginalised posterior mean is closer to the truth for both $\zeta$ and $\mathrm{log}(T_\mathrm{vir})$ for the pspec-only case.
This degree of improvement is roughly in agreement with \citealt{Shimabukuro2016} who perform a Fisher analysis using \textsc{\small 21CMFASTv2} in which they consider the sensitivity levels of LOFAR and MWA.
Although, even in the best case scenario of a perfect observation, the degree of improvement is not as extreme as the Fisher analysis suggests.
This is understandable, since the inverse of the Fisher matrix only provides an estimate of smallest achievable credible limit.

\section{The impact of sample variance and instrumental effects}\label{sec:sv}

\subsection{Modelling the sample-variance error}\label{sec:sv}
A major challenge to performing parameter estimation with 21-cm data and simulations is
correctly accounting for sample variance. Even at the level of the power spectrum this is
difficult as the error due to sample variance is dependent on the 21-cm signal itself,
and therefore the model parameters.
This makes it a great challenge to model the sample-variance error using MC simulations as we have here. One would need to effectively sample the full model parameter space (which for the current most complex \cmmc\,model consists of 17 astrophysical parameters, see \citealt{Qin2020})
at each point performing at least several hundred, ideally thousands of simulations with different initial conditions.
This would realistically require the use of a machine-learning interpolation procedure to make this tractable. 
You would also need to decide a-priori how you are going to chop up your lightcone to measure your statistics as a function of redshift (necessary to effectively capture the evolution of the signal with redshift using such summary statistics), or store all the simulations to avoid being locked into any such choice (not a terribly practical option).
It is therefore interesting to consider whether we might be able to approximate the sample-variance error using an analytic approach.

\begin{figure}
  \centering
    $\renewcommand{\arraystretch}{-0.75}
    \begin{array}{c}
      \includegraphics[trim=0.8cm 0.8cm 0.0cm 0.0cm, clip=true, scale=0.24]{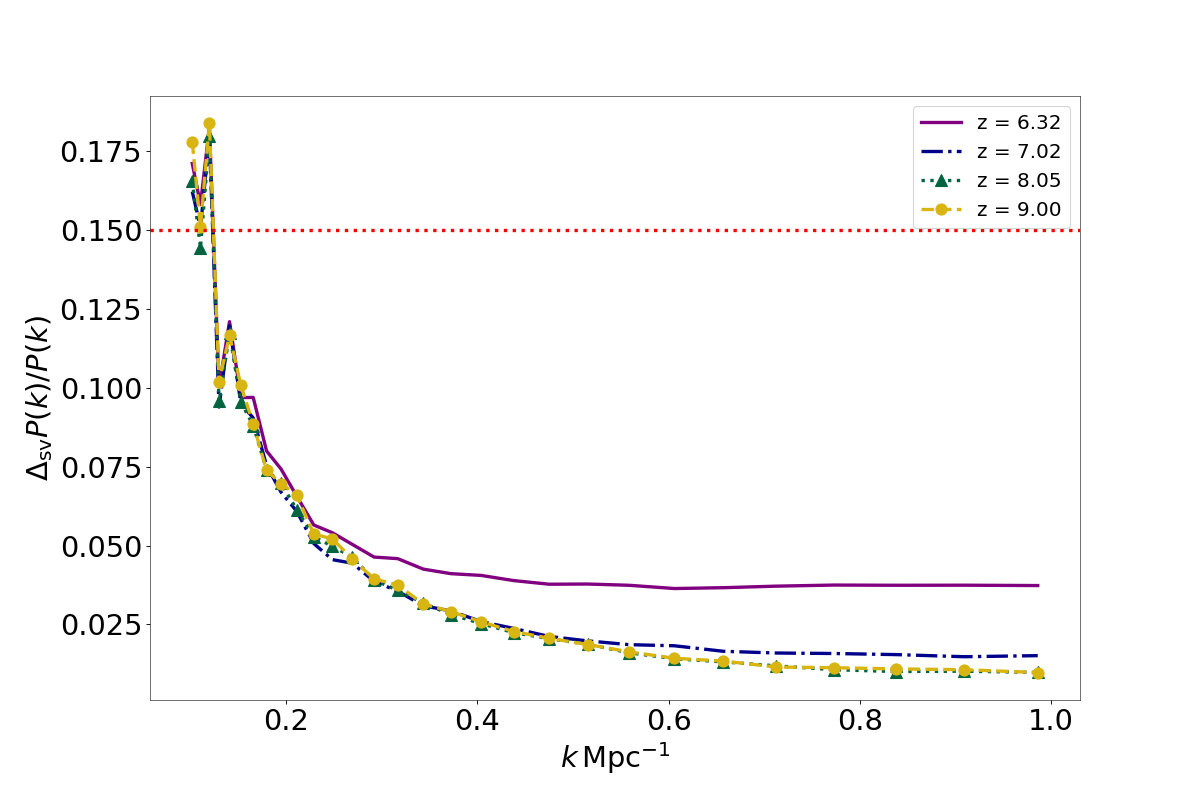}\\
    \end{array}$
  \caption{Plot of the ratio of power-spectrum sample variance (calculated from repeated simulations)
  with the power spectrum for all the redshifts considered.
  We see that simply assuming a fixed sample-variance error of 15\% (marked with the dotted horizontal line) is a reasonable choice and will not result in any biasing of results.}
  \label{fig:sverr_over_Pk}
\end{figure}

We first consider whether using a constant sample variance error could suffice, which would allow us to use a simple factorial error term similar to the modelling error that is built into \cmmc\,(intended to account for numerical inaccuracies in the simulation code).
Under this ansatz, the sample-variance error on the power spectrum of a particular $k$ bin, would be described as $\Delta_\mathrm{sv} P(k) = A\,P(k)$
where $A$ is the error factor; by default $A = 0.15$ is used for the similar but distinct modelling error in \cmmc.
We can estimate the true sample variance using a Monte-carlo approach in which we vary the initial-condition's random seed and random-noise realisation assuming the fiducial model parameters.
If we look at the ratio of this `true' sample-variance error for the power spectrum
with the power spectrum magnitude,
i.e. $\Delta_\mathrm{sv}P(k)/P(k)$, as shown in Figure \ref{fig:sverr_over_Pk}, we see that $\Delta_\mathrm{sv}P(k)/P(k)\lessapprox 0.15$ for all scales and redshifts considered.
Therefore this is a reasonable choice for our fiducial model and we will not get biased results as a result of this assumption.
It is important to emphasise that this may not be universally true for all models,
as such further studies would be necessary to be able to investigate this point more deeply.

\begin{figure}
  \centering
    $\renewcommand{\arraystretch}{-0.75}
    \begin{array}{c}
      \includegraphics[trim=0.7cm 3cm 3.75cm 2.75cm, clip=true, scale=0.23]{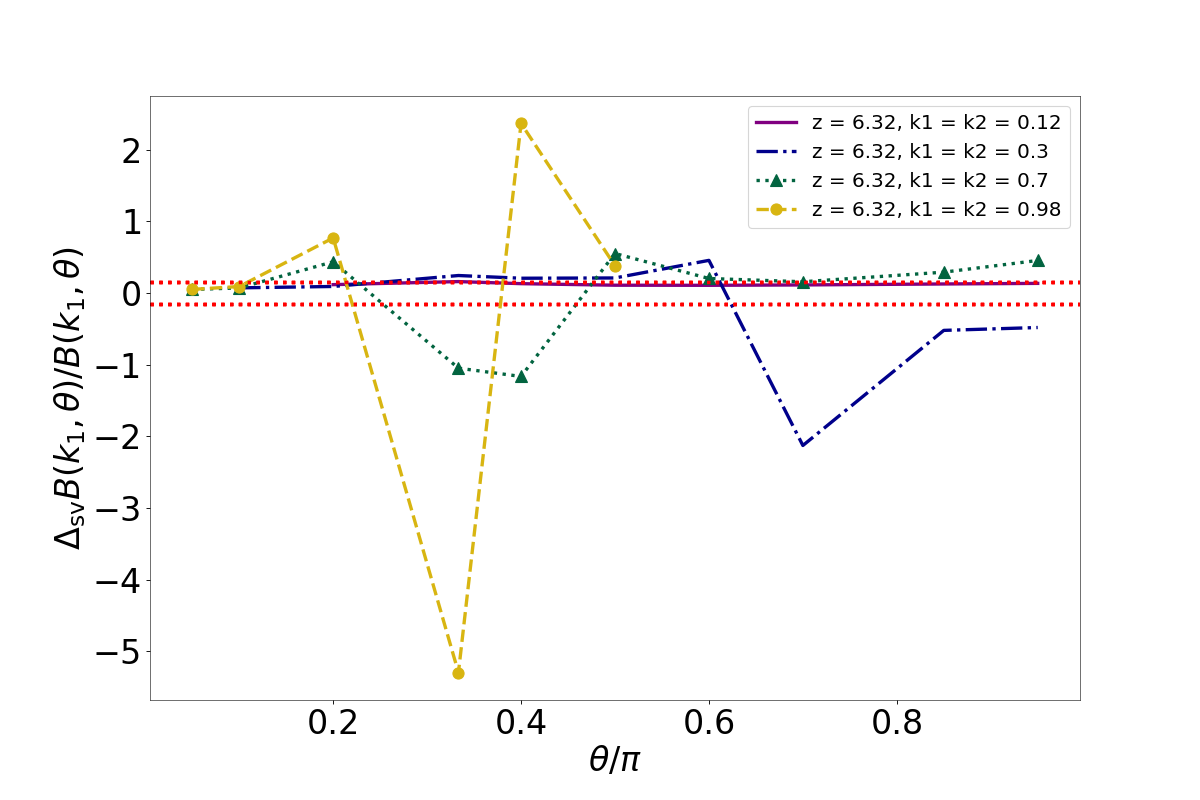}\\
      \includegraphics[trim=0.7cm 3cm 3.75cm 2.75cm, clip=true, scale=0.23]{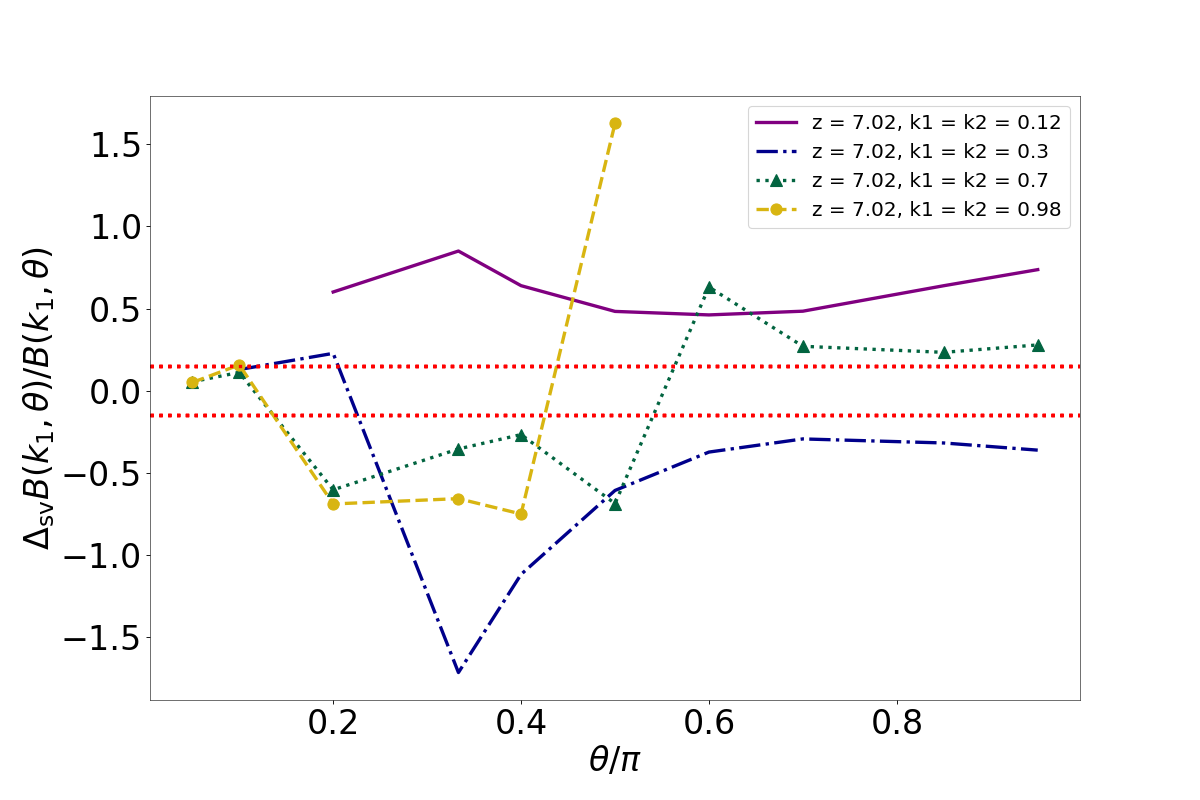}\\
      \includegraphics[trim=0.7cm 3cm 3.75cm 2.75cm, clip=true, scale=0.23]{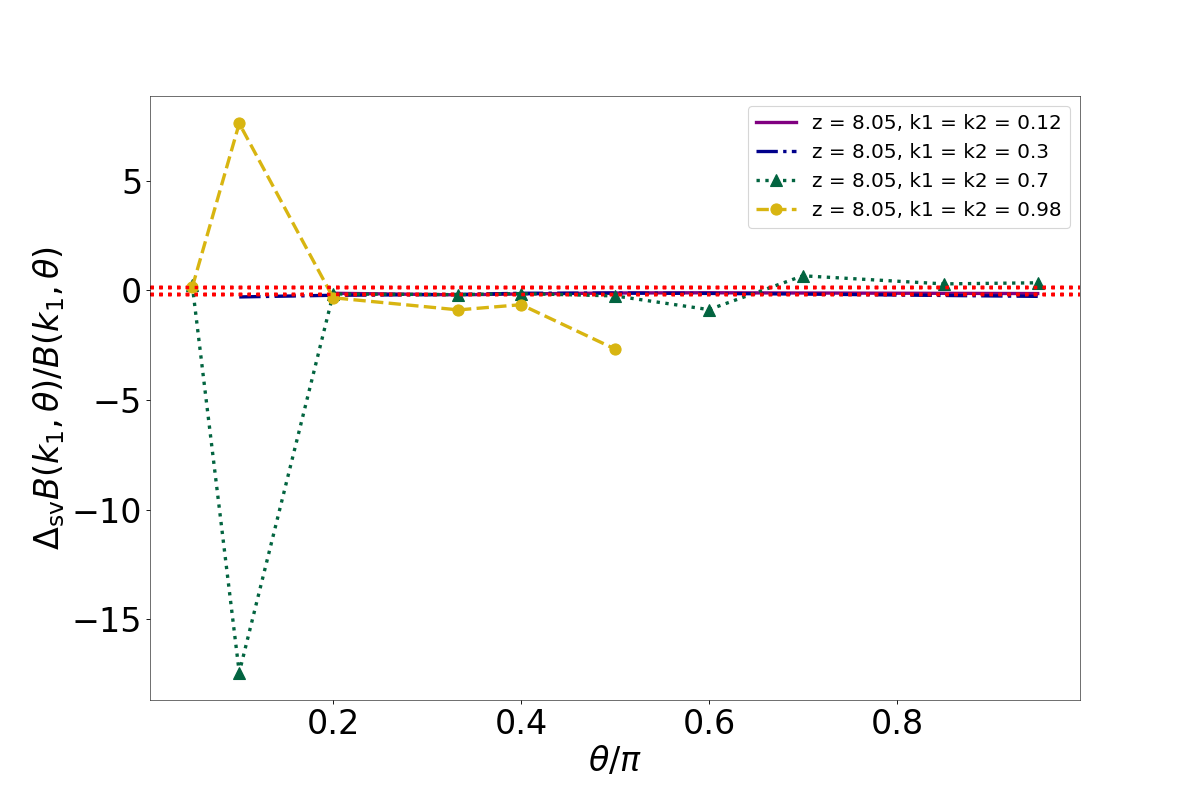}\\
      \includegraphics[trim=0.7cm 0.75cm 3.75cm 2.75cm, clip=true, scale=0.23]{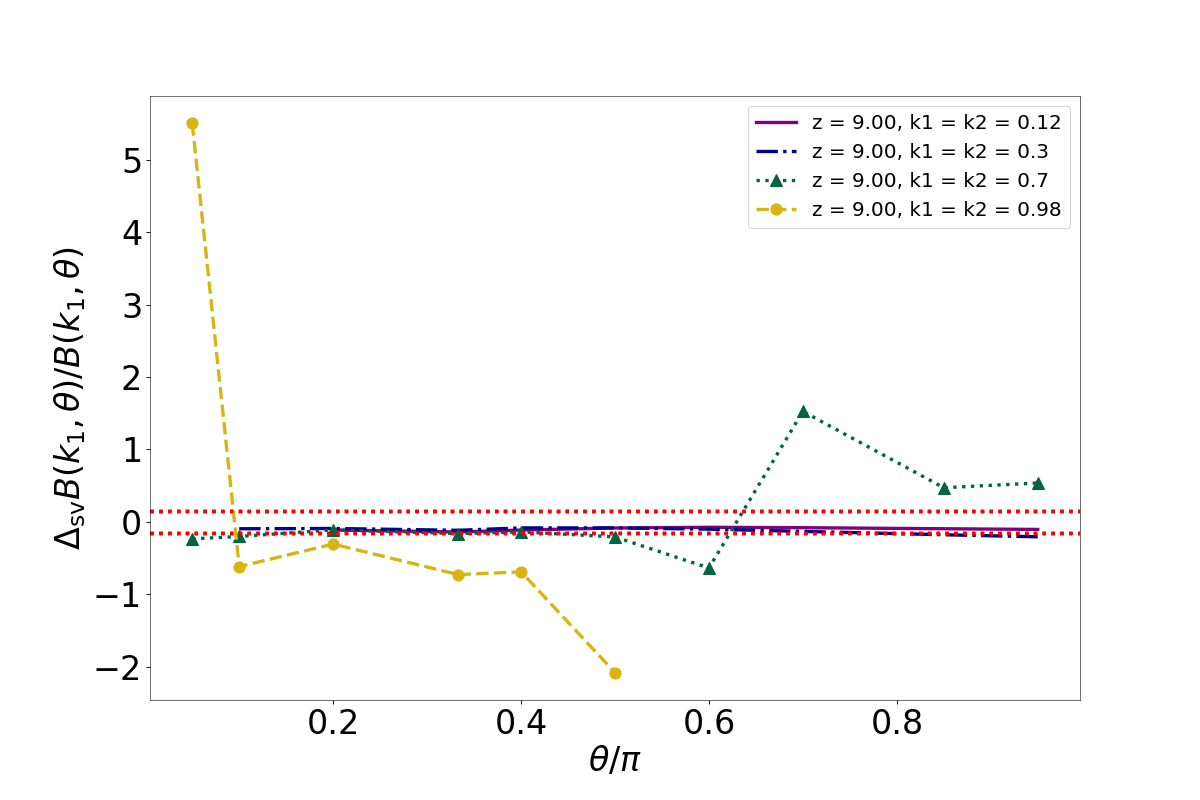}\\
    \end{array}$
  \caption{Plot of the ratio of the isosceles bispectrum sample variance (calculated from repeated simulations)
  and the bispectrum for all the redshifts and triangle configurations considered.
  We see that simply assuming a fixed sample-variance error of 15\% would massively underestimate
  the sample variance for most scales and redshifts considered.}
  \label{fig:sverr_over_BS}
\end{figure}

If we perform the same exercise for the isosceles bispectrum and plot $\Delta_\mathrm{sv}B(k_1, \theta_{12})/B(k_1, \theta_{12})$
as we have in Figure \ref{fig:sverr_over_BS} (solid lines) we see that to assume a sample-variance error of $0.15\,B(k_1, \theta_{12})$ would severely underestimate the sample variance and would undoubtedly impact on our results.
It is also clear that assuming any value for the constant sample-variance error cannot provide a decent approximation to the true sample-variance error calculated using an MC approach.

Assuming the signal is Gaussian, an estimate for the power spectrum sample-variance error is given by
$\Delta^2_{\mathrm{\textsc{sv}}}(k)  = \Delta^2_{21}(k) = k^3/(2\pi^2)P_{21}(k)/\sqrt{N(k)}$, where $P_{21}(k)$ is the 21-cm brightness-temperature power spectrum and $\sqrt{N(k)}$ is the number of times a particular mode has been sampled.
Similarly, we can calculate the theoretical bispectrum sample variance error assuming it is Gaussian distributed (as is often done in the case of Gaussian noise) as,
\begin{equation}
\begin{split}
\left[\Delta_{\mathrm{sv}}B(k_1, k_2, k_3)\right]^2 = 
k_f^3\,\frac{n_{123}}{V_{123}}\,\Delta_\mathrm{sv}P(k_1)\,\Delta_\mathrm{sv}P(k_2)\,\Delta_\mathrm{sv}P(k_3)\,,\\
\end{split}\label{eqn:SVerr}
\end{equation}
in this expression $k_{\rm f}=2\,\pi/L$ is the fundamental $k$ scale, $V_{123} \approx 8.0\pi^2\,k_1\,k_2\,k_3\,(s\,k_{\rm f})^3$
gives the number of fundamental triangles in units of $k_{\rm f}^3$,
$s\,k_{\rm f}$ is the binwidth, and $n_{123} = 1, 2, 6$ for general,
isosceles and equilateral triangle configurations respectively \citep{Scoccimarro1998a, Scoccimarro2004, Liguori2010}.
We assume $s=1$ to obtain the maximum possible estimate for the theoretical sample-variance contribution to the bispectrum using this approximation.

\begin{figure}
  \centering
    $\renewcommand{\arraystretch}{-0.75}
    \begin{array}{c}
      \includegraphics[trim=1.5cm 1.15cm 3.75cm 2.75cm, clip=true, scale=0.25]{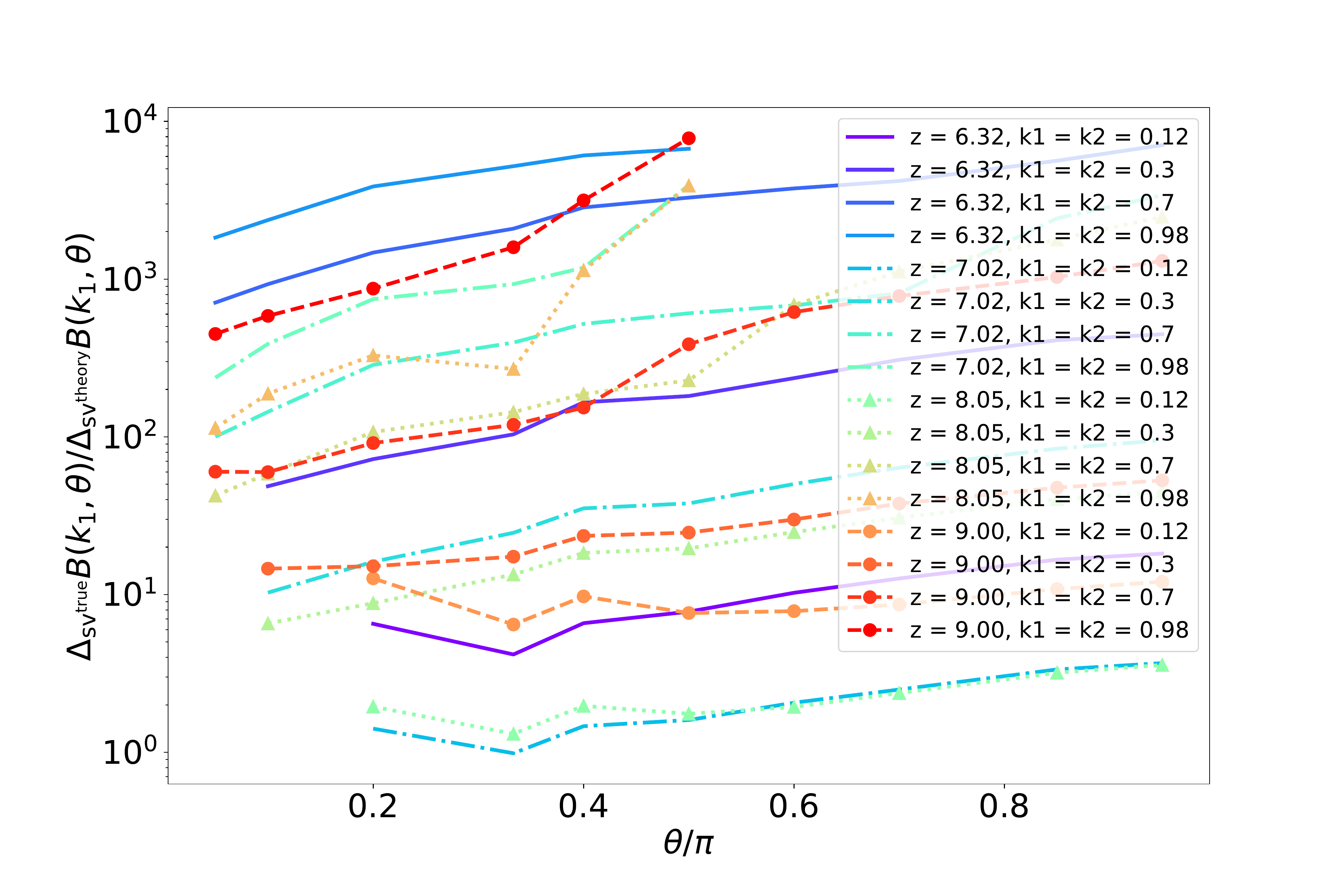}\\
      \end{array}$
      \caption{Ratio of the sample variance on the bispectrum as measured from brute force
      repeat simulation to that measured from theory assuming the signal is Gaussian.
      Solid line correspond to $z=6.3$, dot-dashed to $z=7$, dotted with triangles to $z=8$ and dashed with circles to $z=9$.
      The Gaussian assumption for the sample variance is unable to even qualitatively capture
      the features we see in the simulated sample variance.}
\label{fig:MCerr_over_sverr}
\end{figure}

The ratio of the MC sample-variance error to that calculated using Equation~\ref{eqn:SVerr} is plotted in Figure \ref{fig:MCerr_over_sverr} (where solid line correspond to $z=6.3$, dot-dashed to $z=7$, dotted with triangles to $z=8$ and dashed with circles to $z=9$).
It is clear that
this approximation is orders of magnitude lower than the true sample variance for this box size and resolution.
It is also clear there is no clean connection between this theoretical sample variance and the true sample variance.
It is possible to improve on this theoretical approximation; for example, one can add the trispectrum contribution to the sample variance of the power spectrum as per \citealt{Shaw2020}, which includes a contribution from the non-Gaussianity of the data. We defer such extentions to this aspect of this study to future work.

However, as is evident from the results of Section \ref{sec:ideal}, Figure \ref{fig:MCerr_over_sverr} and the many works studying
non-Gaussianity of the 21-cm signal, our signal is far from Gaussian.
Various works including \citealt{Mondal2016a}, \citealt{Shaw2019}, and \citealt{Shaw2020}
have shown that correctly accounting for this non-Gaussianity in the power-spectrum covariance will have a non-negligible  effect on the resulting parameter constraints (provided large-scale measurements are limited by thermal noise).
Therefore, for the rest of the paper we will use the MC estimated error estimates.

\subsection{Parameter constraints using Monte-Carlo simulated error term}\label{sec:mainresults}
\begin{figure}
  \centering
    $\renewcommand{\arraystretch}{-0.75}
    \begin{array}{c}
      \begin{overpic}[trim=1.0cm 3cm 3.75cm 2.87cm, clip=true, scale=0.225]{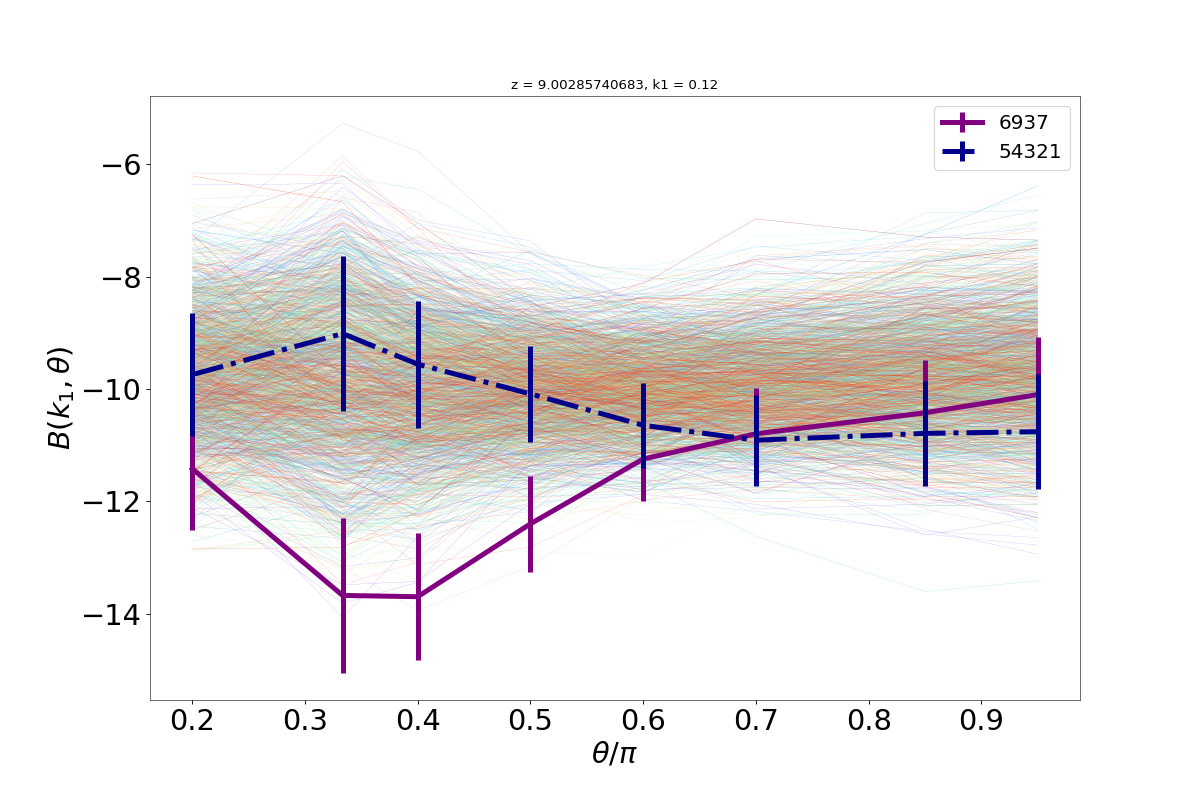}\put(27,115){$k_1=0.12, z=9.0$}
      \end{overpic}\\
      \begin{overpic}[trim=1.0cm 3cm 3.75cm 2.86cm, clip=true, scale=0.225]{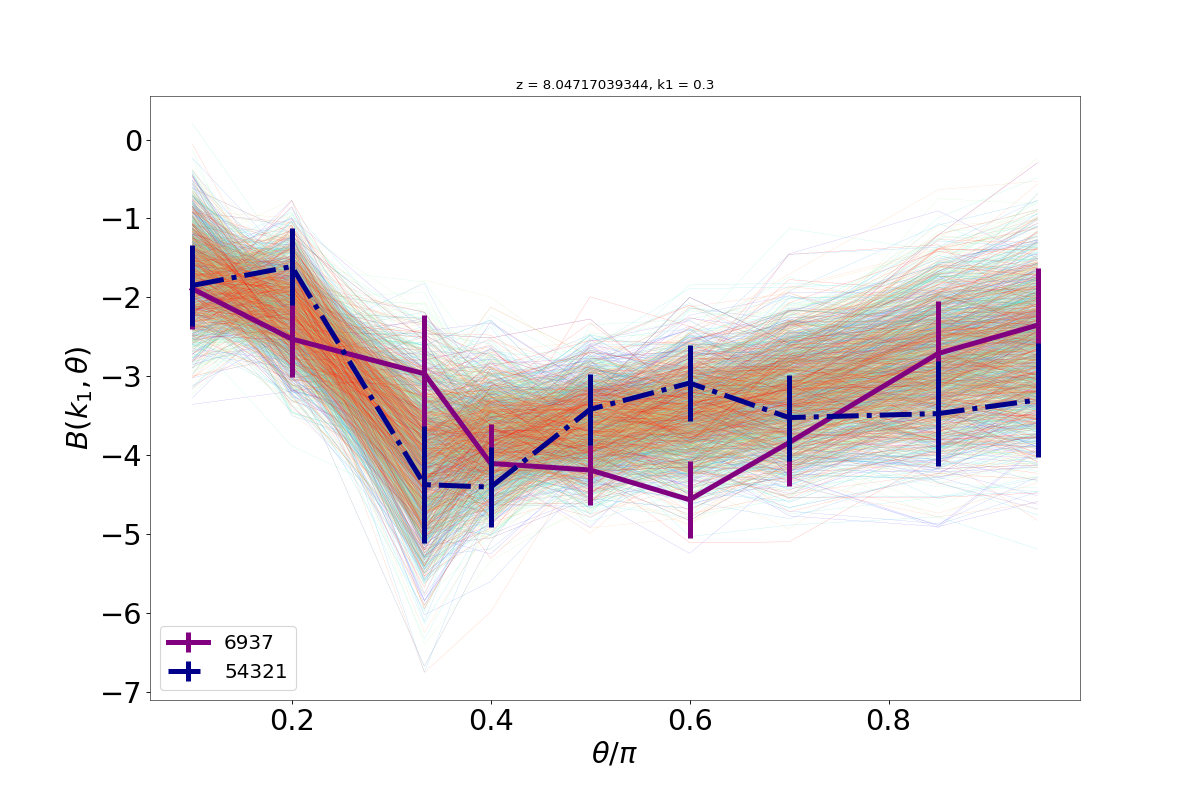}\put(27,115){$k_1=0.3, z=8.0$}
      \end{overpic}\\
      \begin{overpic}[trim=1.0cm 3cm 3.75cm 2.85cm, clip=true, scale=0.225]{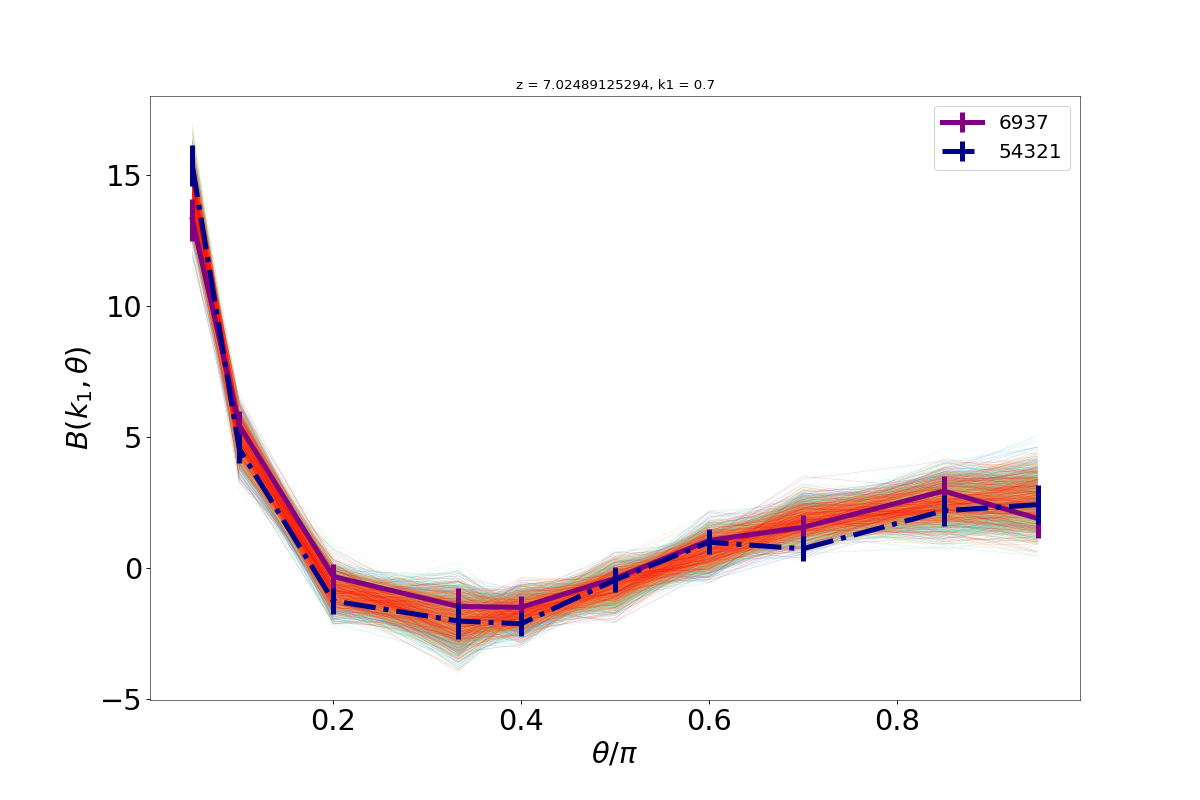}\put(27,115){$k_1=0.7, z=7.0$}
      \end{overpic}\\
      \begin{overpic}[trim=1.0cm 0.75cm 3.75cm 2.84cm, clip=true, scale=0.225]{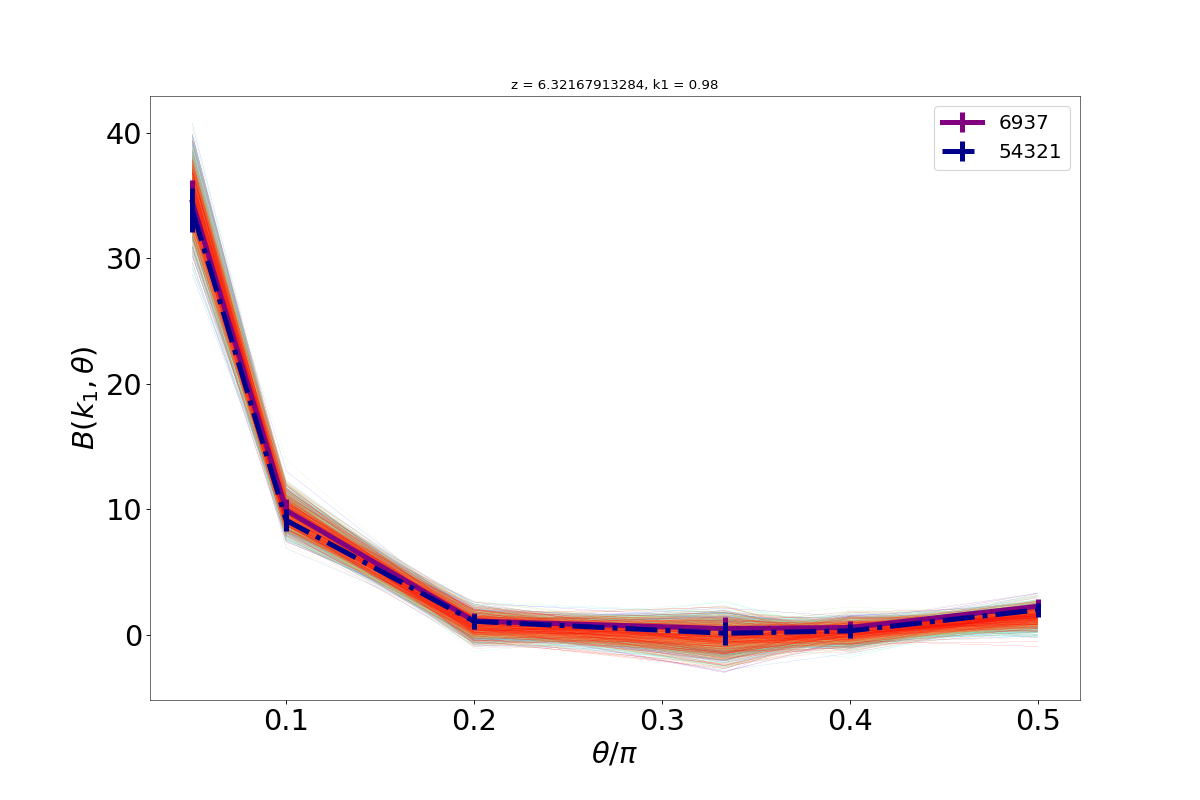}\put(27,130){$k_1=0.98, z=6.32$}
      \end{overpic}\\
    \end{array}$
  \caption{Here we plot with thin lines all 2000 bispectra used in estimating the error due to sample variance
  for our simulation dimensions.
  The plots from top to bottom correspond to $k_1 = [0.12, 0.3, 0.7, 0.98]$ and $z = [9.0, 8.0, 7.0, 6.3]$.
  We overplot the two random seeds used in our parameter estimation analysis chosen from about 50 trial runs to minimise (54321) and maximise (6937) the reduced $\chi^2$
  between them and the mean of the distribution of the thin lines shown by the thin lines in the plot.
 }
  \label{fig:random_seeds}
\end{figure}

The initial conditions of our Universe can impact the outcome of our parameter estimation. To quantify this, we choose a "standard" and an "extreme" model for our mock observations used for parameter inference.  Specifically, we use two different random seeds that exhibit minimal and maximal $\chi^2$ from the mean of the signal, selected from among $\sim$50 different realizations.
In the analysis of this section we use the MCMC estimated noise+sample variance error, but since we are using \cmfast\,for generating our mock observations, we set the modelling error factor to $A=0.0$.
We show the bispectrum of these two random seeds in Figure~\ref{fig:random_seeds}, we also plot
in thin lines the full range of bispectrum produced in the repeat sampling we used to estimate
the 1$\sigma$ sample-variance errors (which are the error bars on each of our random seed bispectra).
The plots from top to bottom correspond to $k_1 = [0.12, 0.3, 0.7, 0.98]$ and $z = [9.0, 8.0, 7.0, 6.3]$.
As can be seen from this plot, seed 6937 is our "extreme" seed and seed 54321 is our "standard" seed.
For the interested reader we have include the equivalent plots for the power spectrum in Appendix \ref{appendix:sv_pspec}.

\begin{figure}
  \centering
    $\renewcommand{\arraystretch}{-0.75}
    \begin{array}{c}
      \includegraphics[trim=0.0cm -0.5cm -0.6cm 0.0cm, clip=true, scale=0.5]{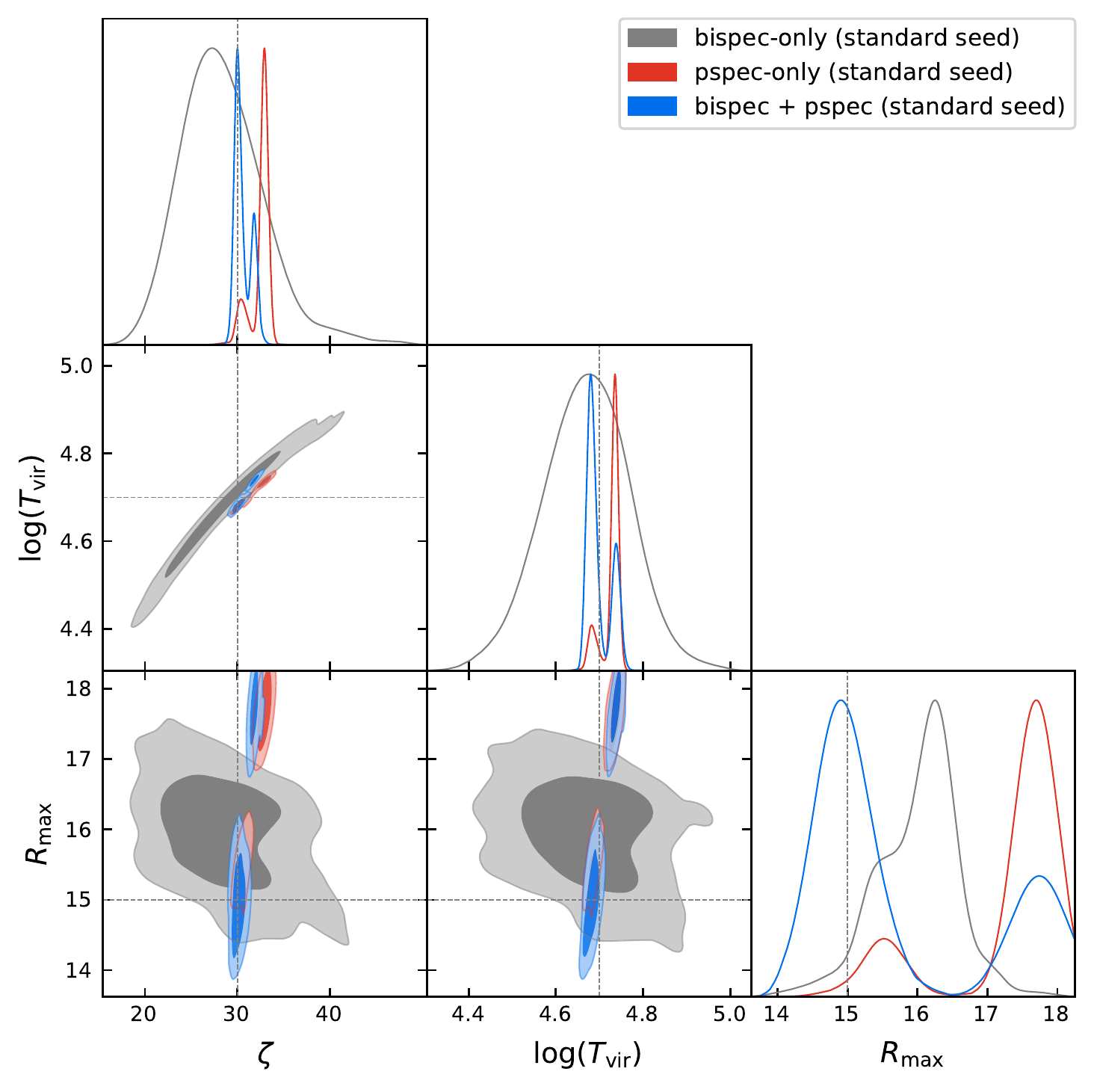}\\
      \includegraphics[trim=1.5cm 0.0cm 0.0cm 3cm, clip=true, scale=0.275]{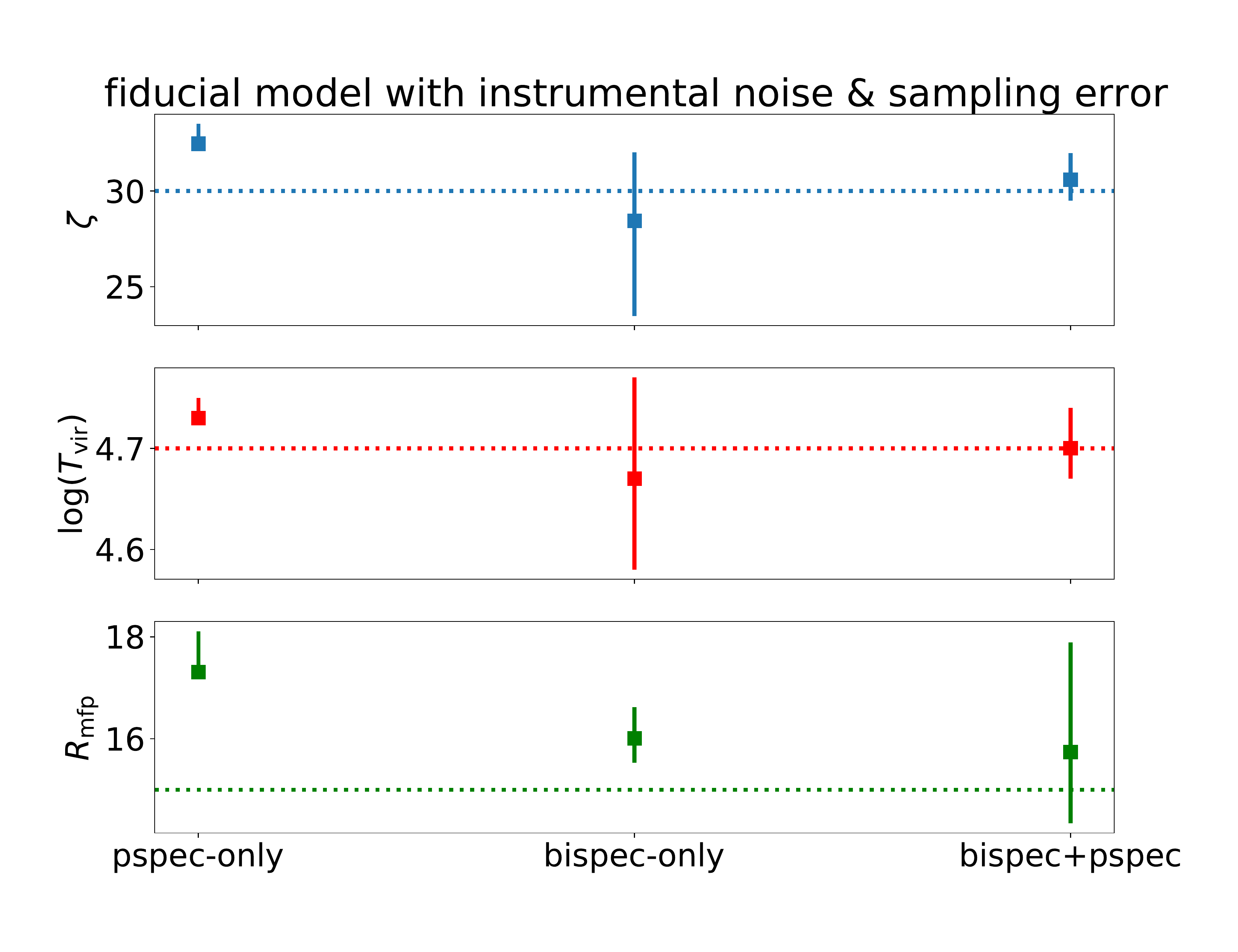}\\
        \end{array}$
  \caption{Corner plot (top) of credible intervals when using mock observed data generated using the fiducial model and the standard seed and the bispec-only (grey contours), the pspec-only (red contours), and bipsec+pspec (blue contours) as summary statistics in the likelihood.
  The blacked dashed lines indicate the parameter values used to generate the mock observed datasets for each model.
  The bottom plot shows the mean +/- 68\% credible intervals for each parameter.
  All include the effects of SKA-LOW (phase 1) $uv$ sampling and noise,
  as well as sample variance, which we model the associated standard deviation using MC methods and using the parameters of the fiducial model.
  Whilst all cases contain the truth within their 95\% credible intervals, the posterior probability mass for the pspec-only case is concentrated in a different region of model parameter space, resulting in biased marginal statistics.
  }
  \label{fig:3param_std_seed}
\end{figure}

The top plot of Figure~\ref{fig:3param_std_seed} shows the resulting credible intervals when we use the standard seed and assume the parameters of our fiducial model for our mock observed data.
The forward model and mock observed data used for the analysis behind this plot both include instrumental effects (i.e. $uv$ sampling and noise).
As before, the largest grey contour shows the bispectrum-only case, the red contours the power-spectrum only case, and the blue contours the bispectrum + power spectrum case.
For both models the true parameters values (marked with the black dashed lines) lie within the 95\% credible intervals for all three combinations of statistic, however for the fiducial model the power spectrum posterior is bimodal.
Furthermore, there is more probability density in the mode that is centred around different parameter values to the truth, leading to biased marginal statistics (this can be seen from the marginalised statistics for this case which we show in the bottom plot of Figure~\ref{fig:3param_std_seed}).
The posterior for the bispectrum-only case has its probability density focused around the true parameter values for $\Tvir$ and $\zeta$, but as with the power spectrum exhibits bias towards larger $\Rmax$.
All is saved by combining the power spectrum with the bispectrum, the marginal statistics of which do not suffer from bias on the inferred parameter values.

\begin{figure}
  \centering
    $\renewcommand{\arraystretch}{-0.75}
    \begin{array}{c}
      \includegraphics[trim=0.0cm -0.5cm -0.6cm 0.0cm, clip=true, scale=0.5]{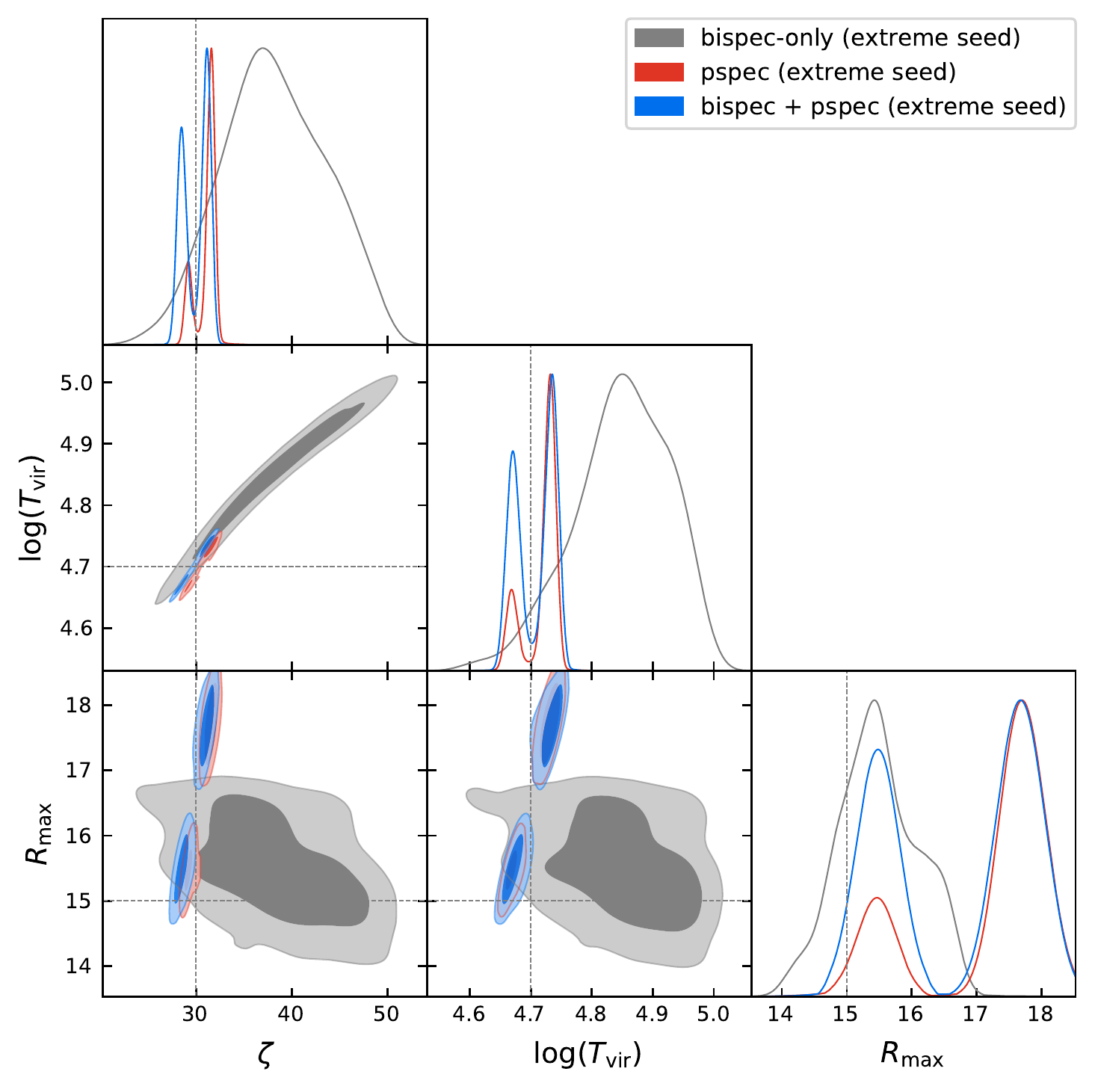}\\
      \includegraphics[trim=1.5cm 0.0cm 0.0cm 3cm, clip=true, scale=0.275]{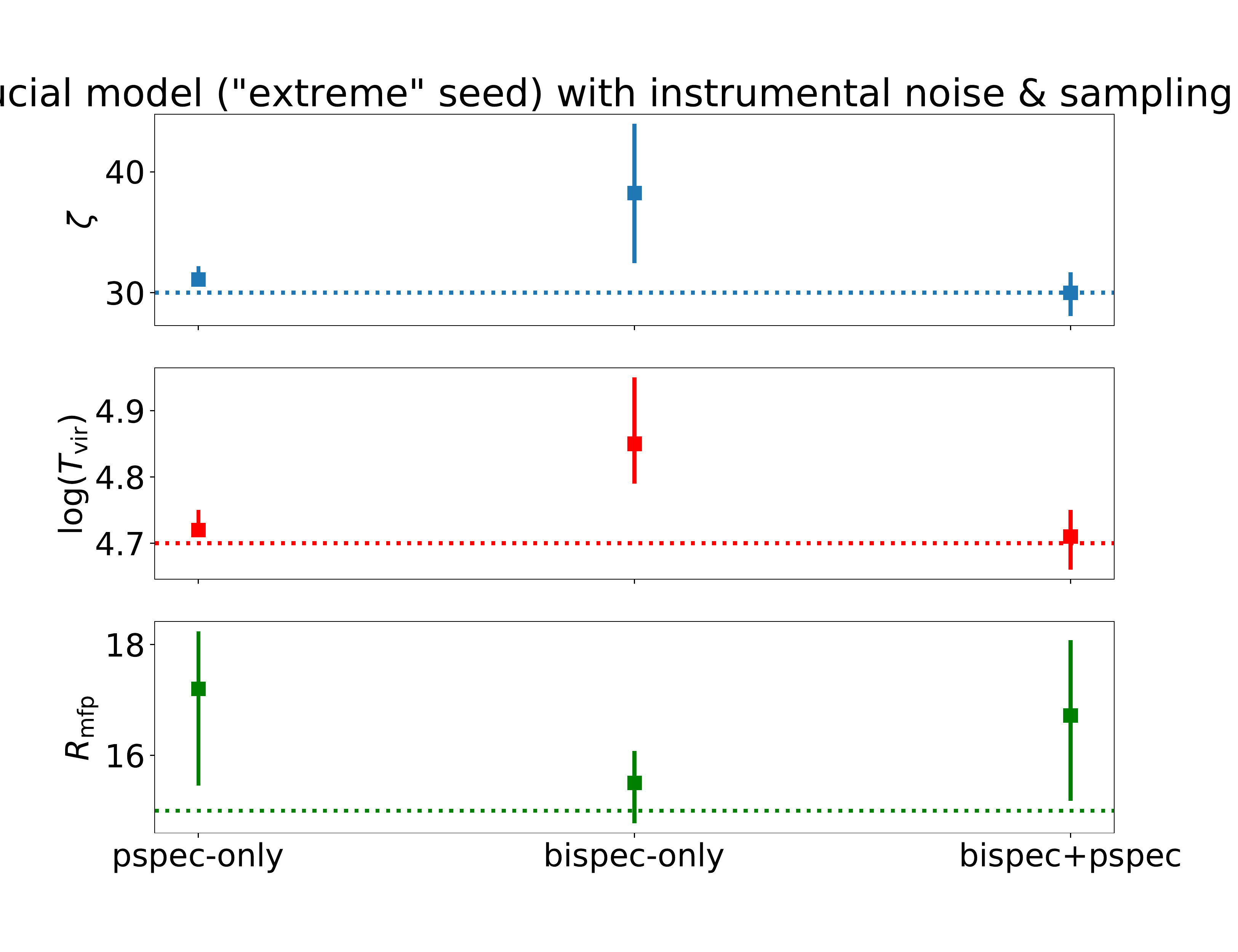}\\
        \end{array}$
  \caption{Corner plot (top) of credible intervals when using mock observed data generated using the fiducial model and the "extreme" seed for the bispec-only (grey contours), pspec-only (red contours), and bipsec+pspec (blue contours) as summary statistics in the likelihood.
  The blacked dashed lines indicate the parameter values used to generate the mock observed datasets for each model.
  The bottom plot shows the mean +/- 68\% credible intervals for each parameter.
  All include SKA-LOW (phase 1) instrumental effects (assuming negligible primary beam effects) as well as sample variance, which we model the associated standard deviation using MC methods and using the parameters of the fiducial model.
  All cases contain the truth within their 95\% credible intervals, albeit in a lower probability region of the posterior.
  }
  \label{fig:3param_extreme_seed}
\end{figure}

If we now consider the results when we use the "extreme" seed for generating our mock observed datasets, then we see that the 95\% credible intervals for all combinations of summary statistic still contain the true model parameters for all parameters. However, they are in a lower probability region of the posterior than they were for the case of the more standard seed.
This can be seen in Figure~\ref{fig:3param_extreme_seed} where the top plot shows the corner plot for the fiducial model with extreme seed.
We see that for the case of the "extreme" seed, the weight is more evenly spread across the two posterior modes resulting in marginal statistics (which are summarised in the bottom plot of Figure~\ref{fig:3param_extreme_seed}) that are less biased than one might imagine from examining the credible intervals by eye.
The bias of the marginal posterior's mean is even reduced for the pspec-only case relative to the results using the more standard seed for the mock observed dataset.
Combining the bispectrum still improves the robustness of the results;
however, the bias on the marginal statistics of $\Rmax$ is not as reduced when the bispectrum and power spectrum are combined as it is for the more standard seed.

\begin{figure}
  \centering
    $\renewcommand{\arraystretch}{-0.75}
    \begin{array}{c}
      \includegraphics[trim=0.0cm -0.5cm -0.6cm 0.0cm, clip=true, scale=0.5]{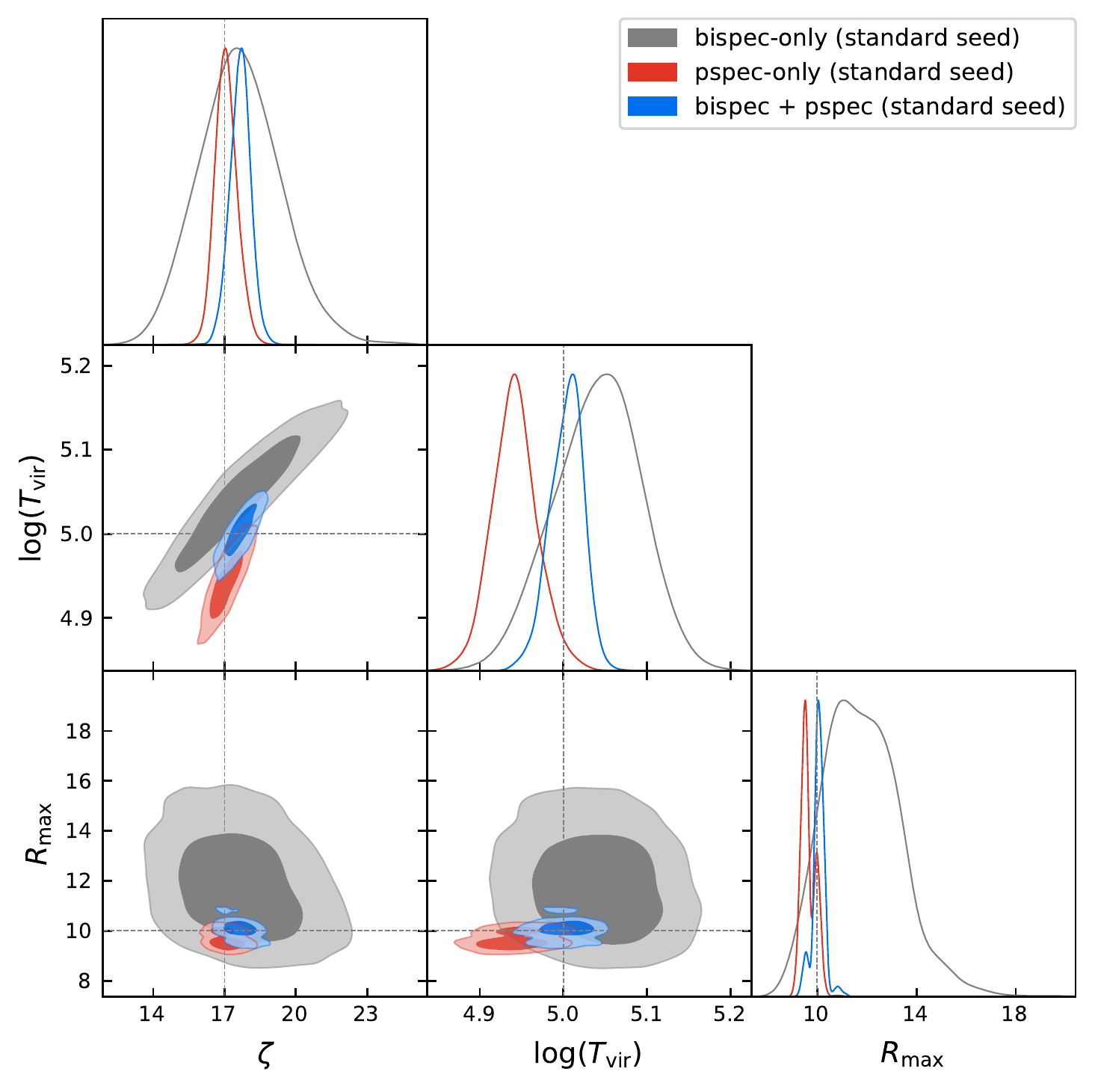}\\
      \includegraphics[trim=1.5cm 0.0cm 0.0cm 3cm, clip=true, scale=0.275]{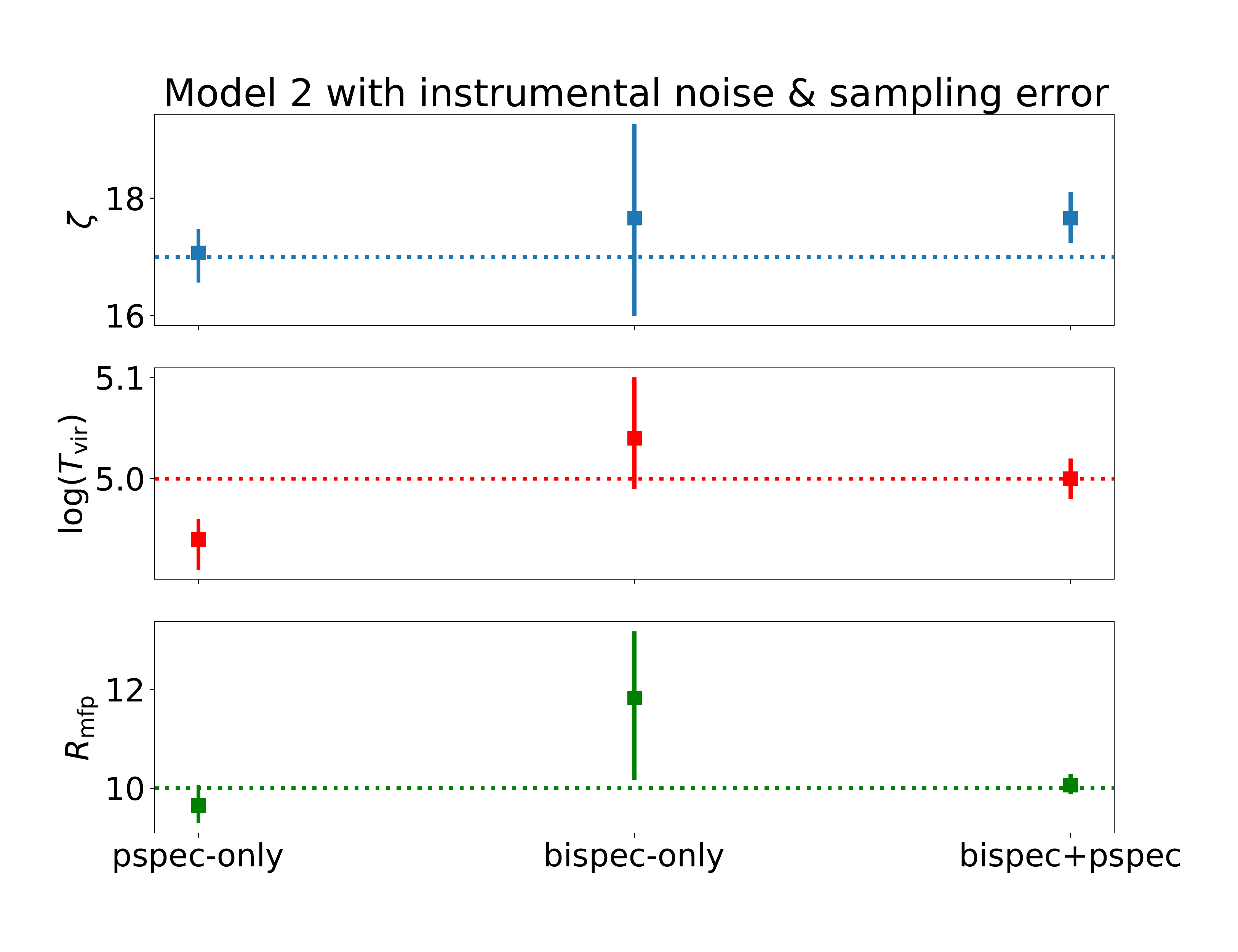}\\
        \end{array}$
  \caption{Corner plot (top) of credible intervals when using mock observed data generated using our late reionization model and the standard seed for the bispec-only (grey contours), the pspec-only (red contours), and bipsec+pspec (blue contours) as summary statistics in the likelihood.
  The blacked dashed lines indicate the parameter values used to generate the mock observed datasets for each model.
  The bottom plot shows the mean +/- 68\% credible intervals for each parameter.
  All include SKA-LOW (phase 1) instrumental effects (assuming negligible primary beam effects) as well as sample variance, which we model the associated standard deviation using MC methods and using the parameters of the fiducial model.
  Whilst all cases contain the truth within their 95\% credible intervals, the posterior probability mass for the pspec-only case is concentrated in a different region of model parameter space, resulting in biased marginal statistics.
  }
  \label{fig:3param_std_seed_m2}
\end{figure}

As can be seen in the corner plot of Figure~\ref{fig:3param_std_seed_m2} (top), there is much less of an issue with bi-modality in the posterior for mock observed data generated with the parameters of our late reionization model; clearly this region of parameter space is less generic (i.e. the model statistics are very distinct from those of other models).
The marginal statistics for this model are summarised in the bottom plot of \ref{fig:3param_std_seed_m2}. 
We see that for our late reionization model using the bispectrum in combination with the power spectrum still overall reduces bias on the marginal statistics (although at the cost of introducing a small bias on the marginal statistics of $\zeta)$ and shrinks the credible intervals relative to those of either statistic alone.
We found that even in test runs where we fixed the modelled initial conditions using the standard seed and the extreme seeds for the data that the results for our late reionization model were still robust with no serious issue with biased results.

\begin{figure}
  \centering
    $\renewcommand{\arraystretch}{-0.75}
    \begin{array}{c}
      \includegraphics[trim=0.0cm -0.5cm -0.6cm 0.0cm, clip=true, scale=0.5]{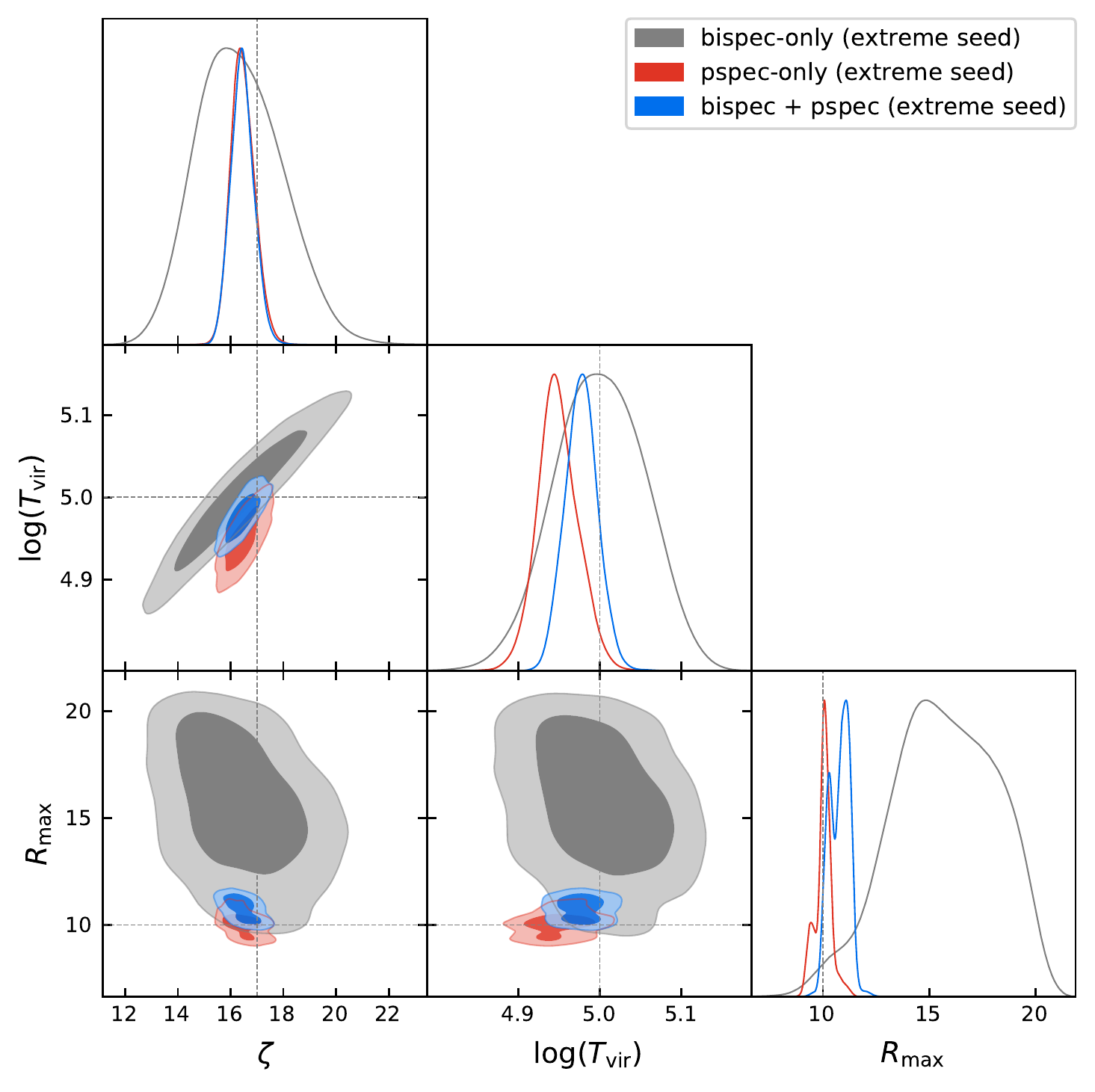}\\
      \includegraphics[trim=1.5cm 0.0cm 0.0cm 3cm, clip=true, scale=0.275]{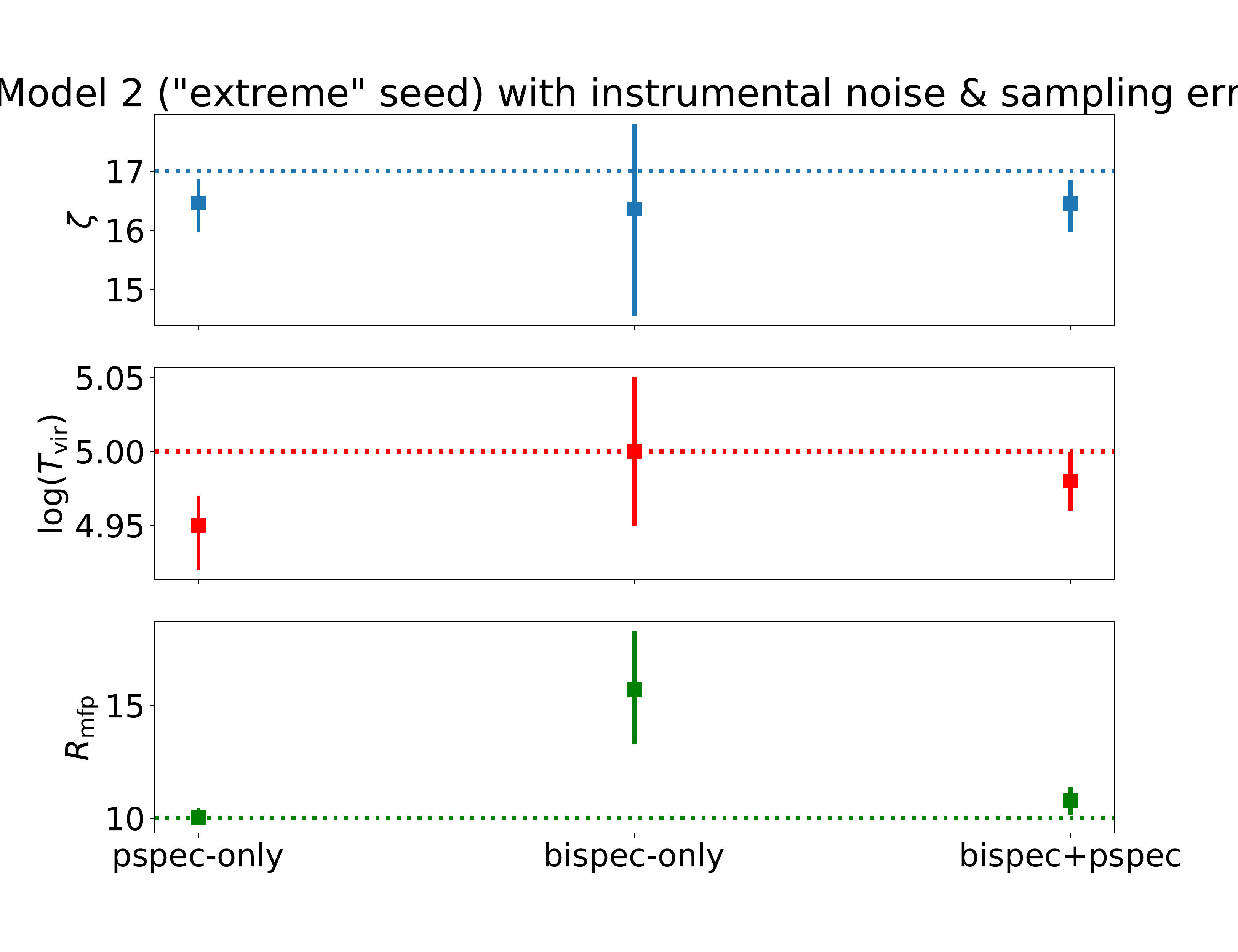}\\
        \end{array}$
  \caption{Corner plot (top) of credible intervals when using mock observed data generated using our late reionization model and the "extreme" seed for the bispec-only (grey contours), pspec-only (red contours), and bipsec+pspec (blue contours) as summary statistics in the likelihood.
  The blacked dashed lines indicate the parameter values used to generate the mock observed datasets for each model.
  The bottom plot shows the mean +/- 68\% credible intervals for each parameter.
  All include SKA-LOW (phase 1) instrumental effects (assuming negligible primary beam effects) as well as sample variance, which we model the associated standard deviation using MC methods and using the parameters of the fiducial model.
  Here the combining the bispectrum with the power spectrum still helps less in relieving bias in the credible intervals, although the marginal statistics are seen to return reasonable predictions of the true parameters.
  }
  \label{fig:3param_extreme_seed_m2}
\end{figure}

Figure~\ref{fig:3param_extreme_seed_m2} shows the results for our late reionization model when the extreme seed is used to generate the mock observed data.
As with the standard seed, the results for our late reionization model are less biased than they are for the fiducial model with the 95\% credible intervals of all combinations of statistic containing the true model and with the combining of the bispectrum and power spectrum improving the quality of the constraints.
As we will discuss further in the following paragraph, this is because this model is at a much early stage of the reionization process for which differences between seeds are suppressed relative to that of the fiducial model.

What is potentially important about the results of the our late reionization model analysis, is that we have used the standard-deviation due to sample variance as calculated for the fiducial model, rather than calculating it for the our late reionization model parameters, i.e. we have seen no serious negative impact from assuming sample variance is the same in both regions of model parameter space, despite them being very different models.
This is likely because the sample-variance error for the later-reionization model is smaller or similar to that of the fiducial model because as the process of reionization is less advanced (the late-reionizaton's ionized fraction is only 0.7 at the lowest redshift we consider as opposed to 0.3 in the fiducial model).
In the later stages of reionization (in the regime of sparse neutral islands) the amplitude varies more between realisations, as can be seen by the trend further away from the theoretical sample-variance with decreasing redshift in Figure \ref{fig:sverr_over_BS}.
This result implies that one could use the sample-variance from a single well-chosen model for all regions in parameter space.
A better, and still tractable, option would be to sparsely sample the sample-variance error in parameter space and use some form of interpolation to approximate the sample-variance error in other regions of parameter space.
Whether or not this would be a sufficient approximation, and whether this finding extends to the full covariance matrix should be addressed in future work.

\begin{figure}
  \centering
    $\renewcommand{\arraystretch}{-0.75}
    \begin{array}{c}
      \includegraphics[trim=0.0cm 0.0cm 0.0cm 3.25cm, clip=true, scale=0.5]{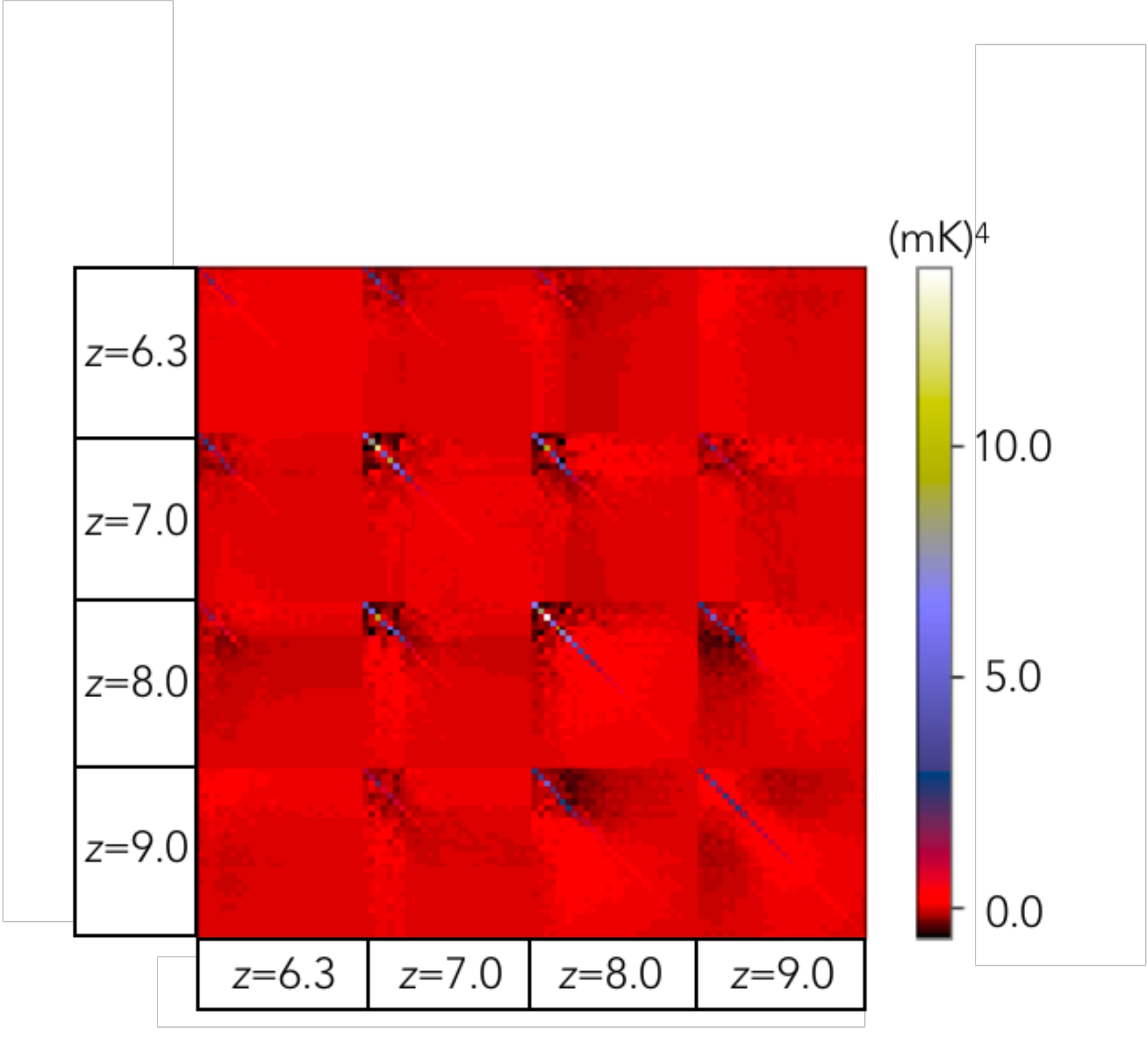}\\
        \end{array}$
  \caption{Covariance matrix for the power spectrum for all bins and redshifts considered here.
  We see that there is correlations between statistical bins in different redshifts, most notably between $z\,=\,7$ and $z\,=\,8$.
  }
  \label{fig:PS_cov}
\end{figure}

It is clear that using a diagonal covariance matrix and assuming independence between statistical bins are not disastrous assumptions in that the true parameters are constrained by the resulting parameter estimation analysis, even if we consider outlier data. However, it will give stronger and more robust results to not make such assumptions and to use a fully multivariate Gaussian likelihood that includes all correlations between the statistical bins, statistics and redshifts.
We have discussed the difficulty of accurately accounting for sample variance errors, it will equally be challenging to capture correlations between redshifts, which can be seen in Figure \ref{fig:PS_cov} where we plot the covariance matrix for the power spectrum.\footnote{We do not plot the correlations between the power spectrum and bispectrum, because the amplitude contribution has been normalised out of our bispectrum analysis; we therefore expect correlations between the two statistics to be negligible.}
These correlations would likely be less severe if we were working with chunks of lightcones, which is the more correct thing to consider; however, it is unlikely that there would be no correlations whatsoever.
It is also likely, that as the complexity of our forward model increases (necessary if we are to fully characterise the instrumental effects, foreground residuals, ionospheric effects, unresolved RFI and polarisation leakage) that assuming a multivariate Gaussian form for the likelihood will be insufficient.

A method to bypass all these issues would be to use likelihood-free inference, which bypasses the need to ever pre-calculate a covariance matrix since the likelihood (or posterior depending on the type of likelihood-free inference) is estimated using forward simulations during the inference process.
It also means one need never explicitly write down a likelihood.
This approach will also be able to deal with cross-correlations of the cosmological signal with the noise and foregrounds biasing parameter-inference results, as seen in \citealt{Nasirudin2020} who perform far more accurate and detailed forward-modelling than that attempted here (they also use a fully multivariate Gaussian likelihood).
We will discuss the application of likelihood-free methods as applied to 21-cm observations in Watkinson, Alsing, Greig \& Mesinger (in prep).

\section{Conclusion} \label{sec:Conc}
In this work, we have added an isosceles bispectrum likelihood module to the established \cmmc\,code that assumes independence between all statistical bins and redshifts.
We are able to make this assumption by using a normalised version of the bispectrum in which the contribution of the power spectrum to the bispectrum has been removed.
In order to perform our analysis we use two new publicly available codes \bifft\,(to measure the bispectrum with sufficient speed) and \pyobs\,(to simulate $uv$ sampling and random instrumental noise for coeval cubes).

We generate various mock observations by varying astrophysical parameters as well as the random seed for initial conditions.
We consider two sets of astrophysical parameters, that result in different reionization histories: a fiducial model and a late reionization model.
We also consider two different random seeds: one chosen to produce relatively standard bispectra (in terms of its $\chi^2$ compared to the mean) and another to produce more extreme outlier bispectra data. 

Various approaches for handling the bispectrum sample-variance error term have also been considered. We find that the bispectrum sample-variance error cannot be effectively described as a percentage of the bispectrum in a given bin, nor by propagating the power spectrum sample-variance error onto the bispectrum under the assumption of Gaussianity.
We find that using the 1$\sigma$ error generated using Monte-Carlo methods for a simple 1D Gaussian likelihood is sufficient to constrain the parameters of the three parameter model of reionization considered here. We also find that using the sample-variance error generated under our fiducial model whilst assuming simulated data from a late-reionization model has no serious negative impact on our results.
This is important as it implies that we may be able to get away with a sparse sampling of the bispectrum sample-variance error as a function of parameter space combined with some form of interpolation to estimate the error term at the unsampled points of parameter space.

We find that combining the power spectra and the bispectrum in the likelihood can significantly reduce the bias away from the input reionization parameters, for all of the mock observations
and models considered here (see also \citealt{Gazagnes2020a}).
For the late-reionization model we also see a reduction in the credible limits.
These findings hold true even if we consider outlier mock observations.

Further work is needed to establish the improvements from using the bispectrum in more complex models for reionization, such as the mass-dependent parametrisation including spin temperature fluctuations \cmfast\,model.
It will also be important for future works that consider the issue of modelling the bispectrum sample variance, to better understand its dependence on simulation resolution and dimensions.
It will also be necessary to get a better understanding of how these results will be impacted by the inclusion of more levels of observational realism, as there has already been indication that foreground residuals and observational effects will be more problematic for the bispectrum \citep{Watkinson2020}.

\section*{Acknowledgements}
CAW’s research is currently supported by a UK Research and Innovation Future Leaders Fellowship, grant number MR/S016066/1 (PI Alkistis Pourtsidou).
CAW also thanks Jonathan Pritchard for financial support during the
early stages of this project via the European Research
Council under ERC grant number 638743-FIRSTDAWN,
as well the ARC Centre for Excellence in All-Sky Astrophysics in 3D Visitor Fund.
This research utilised Queen Mary's Apocrita HPC facility, supported by QMUL ITS Research.
http://doi.org/10.5281/zenodo.438045.
Parts of this research were supported by the Australian Research Council Centre of Excellence for All Sky Astrophysics in 3 Dimensions (ASTRO 3D), through project number CE170100013.
AM acknowledges support from the European Research Council (ERC) under the European Union’s 
Horizon 2020 research and innovation programmes (AIDA -- \#638809).  The results presented here reflect the authors’ views; the ERC is not responsible for their use.

\section{Data availability}
The codes used to produce this work are all publicly available with links to access provided in the main text.

\bibliographystyle{mn2e}
\bibliography{library.bib}

\begin{thebibliography}{}
 \providecommand{\href}[2]{#2}
  \providecommand{\doi}[1]{\href{http://dx.doi.org/#1}{doi:#1}}
  \providecommand{\eprint}[1]{\href{http://arxiv.org/abs/#1}{arXiv:#1}}

\bibitem[\protect\citeauthoryear{Bernardeau, Colombi, Gaztanaga \&
  Scoccimarro}{Bernardeau et~al.}{2001}]{Bernardeau2001}
Bernardeau F.,  Colombi S.,  Gaztanaga E.,    Scoccimarro R.,  2001, Phys.
  Rep., 367, 1, \eprint{0112551}, \doi{10.1016/S0370-1573(02)00135-7}

\bibitem[\protect\citeauthoryear{Brillinger \& Rosenblatt}{Brillinger \&
  Rosenblatt}{1967}]{BrillingerD.R.Rosenblatt1967}
Brillinger D.,  Rosenblatt M.,  1967, in {Bernard Harris} ed., , Spectr. Anal.
  time Ser..
Wiley, New York, pp 189--232

\bibitem[\protect\citeauthoryear{Cram{\'{e}}r}{Cram{\'{e}}r}{1946}]{Cramer1946}
Cram{\'{e}}r H.,  1946, {Mathematical Methods of Statistics (PMS-9)}.
Princeton University Press, \doi{10.1515/9781400883868}

\bibitem[\protect\citeauthoryear{Dewdney}{Dewdney}{2016}]{Dewdney2016}
Dewdney P.,  2016, Technical report, {SKA1 SYSTEM BASELINE DESIGN V2}.
SKA Office

\bibitem[\protect\citeauthoryear{Fisher}{Fisher}{1935}]{Fisher1935}
Fisher R.~A.,  1935, J. R. Stat. Soc., 98, 39, \doi{10.2307/2342435}

\bibitem[\protect\citeauthoryear{Foreman-Mackey, Hogg, Lang \&
  Goodman}{Foreman-Mackey et~al.}{2013}]{Foreman-Mackey2013}
Foreman-Mackey D.,  Hogg D.~W.,  Lang D.,    Goodman J.,  2013, Publ. Astron.
  Soc. Pacific, 125, 306, \eprint{1202.3665}, \doi{10.1086/670067}

\bibitem[\protect\citeauthoryear{Furlanetto \& Oh}{Furlanetto \&
  Oh}{2005}]{Furlanetto2005}
Furlanetto S.~R.,  Oh S.~P.,  2005, MNRAS, 363, 1031, \eprint{0505065},
  \doi{10.1111/j.1365-2966.2005.09505.x}

\bibitem[\protect\citeauthoryear{Furlanetto, Zaldarriaga \&
  Hernquist}{Furlanetto et~al.}{2004}]{Furlanetto2004a}
Furlanetto S.~R.,  Zaldarriaga M.,    Hernquist L.,  2004, Astrophys. J., 613,
  1, \eprint{0403697}, \doi{10.1086/423025}

\bibitem[\protect\citeauthoryear{Gazagnes, Koopmans \& Wilkinson}{Gazagnes
  et~al.}{2020}]{Gazagnes2020a}
Gazagnes S.,  Koopmans L.~V.,    Wilkinson M.~H.,  2020, arXiv,
  \eprint{2011.08260}

\bibitem[\protect\citeauthoryear{Goodman \& Weare}{Goodman \&
  Weare}{2010}]{Goodman2010}
Goodman J.,  Weare J.,  2010, Commun. Appl. Math. Comput. Sci., 5, 65,
  \doi{10.2140/camcos.2010.5.65}

\bibitem[\protect\citeauthoryear{Gorce \& Pritchard}{Gorce \&
  Pritchard}{2019}]{Gorce2019}
Gorce A.,  Pritchard J.~R.,  2019, \eprint{1903.11402}

\bibitem[\protect\citeauthoryear{Greig \& Mesinger}{Greig \&
  Mesinger}{2015a}]{Greig2015a}
Greig B.,  Mesinger A.,  2015a, MNRAS, 449, 4246, \doi{10.1093/mnras/stv571}

\bibitem[\protect\citeauthoryear{Greig \& Mesinger}{Greig \&
  Mesinger}{2017b}]{Greig2017}
Greig B.,  Mesinger A.,  2017b, MNRAS, 472, 2651, \doi{10.1093/mnras/stx2118}

\bibitem[\protect\citeauthoryear{Hinich \& Clay}{Hinich \&
  Clay}{1968}]{Hinich1968}
Hinich M.~J.,  Clay C.~S.,  1968, Rev. Geophys., 6, 347,
  \doi{10.1029/RG006i003p00347}

\bibitem[\protect\citeauthoryear{Hinich \& Messer}{Hinich \&
  Messer}{1995}]{Hinich1995}
Hinich M.,  Messer H.,  1995, IEEE Trans. Signal Process., 43, 2130,
  \doi{10.1109/78.414775}

\bibitem[\protect\citeauthoryear{Hinich \& Wolinsky}{Hinich \&
  Wolinsky}{2005}]{Hinich2005}
Hinich M.~J.,  Wolinsky M.,  2005, J. Stat. Plan. Inference, 130, 405,
  \doi{10.1016/J.JSPI.2003.12.022}

\bibitem[\protect\citeauthoryear{Hutter, Watkinson, Seiler, Dayal, Sinha \&
  Croton}{Hutter et~al.}{2019}]{Hutter2019}
Hutter A.,  Watkinson C.~A.,  Seiler J.,  Dayal P.,  Sinha M.,    Croton D.~J.,
   2019, MNRAS, \doi{10.1093/mnras/stz3139}

\bibitem[\protect\citeauthoryear{Iliev, Mellema, Ahn, Shapiro, Mao \&
  Pen}{Iliev et~al.}{2014}]{Iliev2014}
Iliev I.~T.,  Mellema G.,  Ahn K.,  Shapiro P.~R.,  Mao Y.,    Pen U.-L.,
  2014, MNRAS, p.~725, \eprint{1310.7463}

\bibitem[\protect\citeauthoryear{Kaur, Gillet \& Mesinger}{Kaur
  et~al.}{2020}]{Kaur2020}
Kaur H.~D.,  Gillet N.,    Mesinger A., , 2020, {Minimum size of cosmological
  21-cm simulations}

\bibitem[\protect\citeauthoryear{Kim \& Powers}{Kim \& Powers}{1978}]{Kim1978}
Kim Y.~C.,  Powers E.~J.,  1978, Phys. Fluids, 21, 1452, \doi{10.1063/1.862365}

\bibitem[\protect\citeauthoryear{Koopmans et~al.,}{Koopmans
  et~al.}{2015}]{Koopmans2015}
Koopmans L.  et~al., 2015, Proc. Adv. Astrophys. with Sq. Km. Array, AASKA14

\bibitem[\protect\citeauthoryear{Lewis}{Lewis}{2011}]{Lewis2011}
Lewis A.,  2011, J. Cosmol. Astropart. Phys., 10, 1475, \eprint{1107.5431},
  \doi{10.1088/1475-7516/2011/10/026}

\bibitem[\protect\citeauthoryear{Liguori, Sefusatti, Fergusson \&
  Shellard}{Liguori et~al.}{2010}]{Liguori2010}
Liguori M.,  Sefusatti E.,  Fergusson J.~R.,    Shellard E. P.~S.,  2010, Adv.
  Astron., 2010, 64, \eprint{1001.4707}, \doi{10.1155/2010/980523}

\bibitem[\protect\citeauthoryear{Majumdar, Pritchard, Mondal, Watkinson,
  Bharadwaj \& Mellema}{Majumdar et~al.}{2017}]{Majumdar2017}
Majumdar S.,  Pritchard J.~R.,  Mondal R.,  Watkinson C.~A.,  Bharadwaj S.,
  Mellema G.,  2017, MNRAS, 476, 4007, \eprint{1708.08458},
  \doi{10.1093/mnras/sty535}

\bibitem[\protect\citeauthoryear{Mellema et~al.,}{Mellema
  et~al.}{2013}]{Mellema2013}
Mellema G.  et~al., 2013, Exp. Astron., 36, 235, \eprint{1210.0197}

\bibitem[\protect\citeauthoryear{Mesinger \& Furlanetto}{Mesinger \&
  Furlanetto}{2007}]{Mesinger2007}
Mesinger A.,  Furlanetto S.~R.,  2007, Astrophys. J., 669, 663,
  \doi{10.1086/521806}

\bibitem[\protect\citeauthoryear{Mesinger, Furlanetto \& Cen}{Mesinger
  et~al.}{2010}]{Signal2010}
Mesinger A.,  Furlanetto S.,    Cen R.,  2010, MNRAS, 411, 955,
  \eprint{1003.3878}, \doi{10.1111/j.1365-2966.2010.17731.x}

\bibitem[\protect\citeauthoryear{Mondal, Bharadwaj \& Majumdar}{Mondal
  et~al.}{2016}]{Mondal2016a}
Mondal R.,  Bharadwaj S.,    Majumdar S.,  2016, Mon. Not. R. Astron. Soc. Vol.
  464, Issue 3, p.2992-3004, 464, 2992, \eprint{1606.03874},
  \doi{10.1093/mnras/stw2599}

\bibitem[\protect\citeauthoryear{Murray, Greig, Mesinger, Mu{\~{n}}oz, Qin,
  Park \& Watkinson}{Murray et~al.}{2020}]{Murray2020}
Murray S.,  Greig B.,  Mesinger A.,  Mu{\~{n}}oz J.,  Qin Y.,  Park J.,
  Watkinson C.,  2020, J. Open Source Softw., 5, 2582, \eprint{2010.15121},
  \doi{10.21105/joss.02582}

\bibitem[\protect\citeauthoryear{Nasirudin, Murray, Trott, Greig, Joseph \&
  Power}{Nasirudin et~al.}{2020}]{Nasirudin2020}
Nasirudin A.,  Murray S.~G.,  Trott C.~M.,  Greig B.,  Joseph R.~C.,    Power
  C.,  2020, Astrophys. J., 893, 118, \eprint{2003.08552},
  \doi{10.3847/1538-4357/ab8003}

\bibitem[\protect\citeauthoryear{Park, Mesinger, Greig \& Gillet}{Park
  et~al.}{2018}]{Park2018}
Park J.,  Mesinger A.,  Greig B.,    Gillet N.,  2018, Mon. Not. R. Astron.
  Soc. Vol. 484, Issue 1, p.933-949, 484, 933, \eprint{1809.08995},
  \doi{10.1093/mnras/stz032}

\bibitem[\protect\citeauthoryear{Pober et~al.,}{Pober
  et~al.}{2013a}]{Pober2013}
Pober J.~C.  et~al., 2013a, Astron. J., 145, 65,
  \doi{10.1088/0004-6256/145/3/65}

\bibitem[\protect\citeauthoryear{Pober et~al.,}{Pober
  et~al.}{2014b}]{Pober2014}
Pober J.~C.  et~al., 2014b, Astron. J., 782, 66, \eprint{1310.7031}

\bibitem[\protect\citeauthoryear{Qin, Mesinger, Park, Greig \& Mu{\~{n}}oz}{Qin
  et~al.}{2020}]{Qin2020}
Qin Y.,  Mesinger A.,  Park J.,  Greig B.,    Mu{\~{n}}oz J.~B.,  2020, MNRAS,
  495, 123, \eprint{2003.04442}, \doi{10.1093/mnras/staa1131}

\bibitem[\protect\citeauthoryear{Rao}{Rao}{1945}]{Rao1945}
Rao C.,  1945, Bull. Calcutta Math. Soc., 37, 81

\bibitem[\protect\citeauthoryear{Scoccimarro}{Scoccimarro}{2015}]{Scoccimarro2015}
Scoccimarro R.,  2015, Phys. Rev. D, Vol. 92, Issue 8, id.083532, 92,
  \eprint{1506.02729}, \doi{10.1103/PhysRevD.92.083532}

\bibitem[\protect\citeauthoryear{Scoccimarro, Sefusatti \&
  Zaldarriaga}{Scoccimarro et~al.}{2004a}]{Scoccimarro2004}
Scoccimarro R.,  Sefusatti E.,    Zaldarriaga M.,  2004a, Phys. Rev. D, 69,
  1550, \eprint{0312286}, \doi{10.1103/PhysRevD.69.103513}

\bibitem[\protect\citeauthoryear{Scoccimarro, Colombi, Fry, Frieman, Hivon \&
  Melott}{Scoccimarro et~al.}{1998b}]{Scoccimarro1998a}
Scoccimarro R.,  Colombi S.,  Fry J.~N.,  Frieman J.~A.,  Hivon E.,    Melott
  A.,  1998b, Astrophys. J., 496, 586, \doi{10.1086/305399}

\bibitem[\protect\citeauthoryear{Sefusatti, Crocce, Scoccimarro \&
  Couchman}{Sefusatti et~al.}{2016}]{Sefusatti2015}
Sefusatti E.,  Crocce M.,  Scoccimarro R.,    Couchman H.,  2016, MNRAS, 460,
  3624, \eprint{1512.07295}, \doi{10.1093/mnras/stw1229}

\bibitem[\protect\citeauthoryear{Shaw, Bharadwaj \& Mondal}{Shaw
  et~al.}{2019b}]{Shaw2019}
Shaw A.~K.,  Bharadwaj S.,    Mondal R.,  2019b, \eprint{1902.08706},
  \doi{10.1093/mnras/stz1561}

\bibitem[\protect\citeauthoryear{Shaw, Bharadwaj \& Mondal}{Shaw
  et~al.}{2020a}]{Shaw2020}
Shaw A.~K.,  Bharadwaj S.,    Mondal R.,  2020a, MNRAS, \eprint{2005.06535},
  \doi{10.1093/mnras/staa2090}

\bibitem[\protect\citeauthoryear{Shimabukuro, Yoshiura, Takahashi, Yokoyama \&
  Ichiki}{Shimabukuro et~al.}{2016a}]{Shimabukuro2016}
Shimabukuro H.,  Yoshiura S.,  Takahashi K.,  Yokoyama S.,    Ichiki K.,
  2016a, p.~11, \eprint{1608.00372}

\bibitem[\protect\citeauthoryear{Shimabukuro, Yoshiura, Takahashi, Yokoyama \&
  Ichiki}{Shimabukuro et~al.}{2016b}]{Shimabukuro2016a}
Shimabukuro H.,  Yoshiura S.,  Takahashi K.,  Yokoyama S.,    Ichiki K.,
  2016b, MNRAS, 468, 1542, \eprint{1608.00372}, \doi{10.1093/mnras/stx530}

\bibitem[\protect\citeauthoryear{Sobacchi \& Mesinger}{Sobacchi \&
  Mesinger}{2014}]{Sobacchi2014}
Sobacchi E.,  Mesinger A.,  2014, MNRAS, pp 1662--1673, \eprint{1402.2298}

\bibitem[\protect\citeauthoryear{Tegmark, Taylor \& Heavens}{Tegmark
  et~al.}{1997}]{Tegmark1996}
Tegmark M.,  Taylor A.,    Heavens A.,  1997, Astrophys. Journal, Vol. 480,
  Issue 1, pp. 22-35., 480, 22, \eprint{9603021}, \doi{10.1086/303939}

\bibitem[\protect\citeauthoryear{Trott et~al.,}{Trott et~al.}{2019}]{Trott2019}
Trott C.~M.  et~al., 2019, Publ. Astron. Soc. Aust., 36, e023,
  \eprint{1905.07161}, \doi{10.1017/pasa.2019.15}

\bibitem[\protect\citeauthoryear{Watkinson, Majumdar \& Pritchard}{Watkinson
  et~al.}{2017a}]{Watkinson2017}
Watkinson C.~A.,  Majumdar S.,    Pritchard J.~R.,  2017a, MNRAS, 472, 2436,
  \eprint{1705.06284}, \doi{10.1093/mnras/stx2130}

\bibitem[\protect\citeauthoryear{Watkinson et~al.,}{Watkinson
  et~al.}{2019b}]{Watkinson2018}
Watkinson C.~A.  et~al., 2019b, MNRAS, 482, 2653, \eprint{1808.02372},
  \doi{10.1093/mnras/sty2740}

\bibitem[\protect\citeauthoryear{Watkinson, Trott \& Hothi}{Watkinson
  et~al.}{2021c}]{Watkinson2020}
Watkinson C.~A.,  Trott C.~M.,    Hothi I.,  2021c, MNRAS, 501, 367,
  \doi{10.1093/mnras/staa3677}

\bibitem[\protect\citeauthoryear{Yoshiura, Shimabukuro, Takahashi, Momose,
  Nakanishi \& Imai}{Yoshiura et~al.}{2015}]{Yoshiura2015}
Yoshiura S.,  Shimabukuro H.,  Takahashi K.,  Momose R.,  Nakanishi H.,    Imai
  H.,  2015, MNRAS, 451, 266, \doi{10.1093/mnras/stv855}

\end{thebibliography}

\appendix
\section{The sample variance of the power spectrum}
\label{appendix:sv_pspec}

\begin{figure}
  \centering
    $\renewcommand{\arraystretch}{-0.75}
    \begin{array}{c}
      \begin{overpic}[trim=1.0cm 3cm 3.75cm 2.87cm, clip=true, scale=0.225]{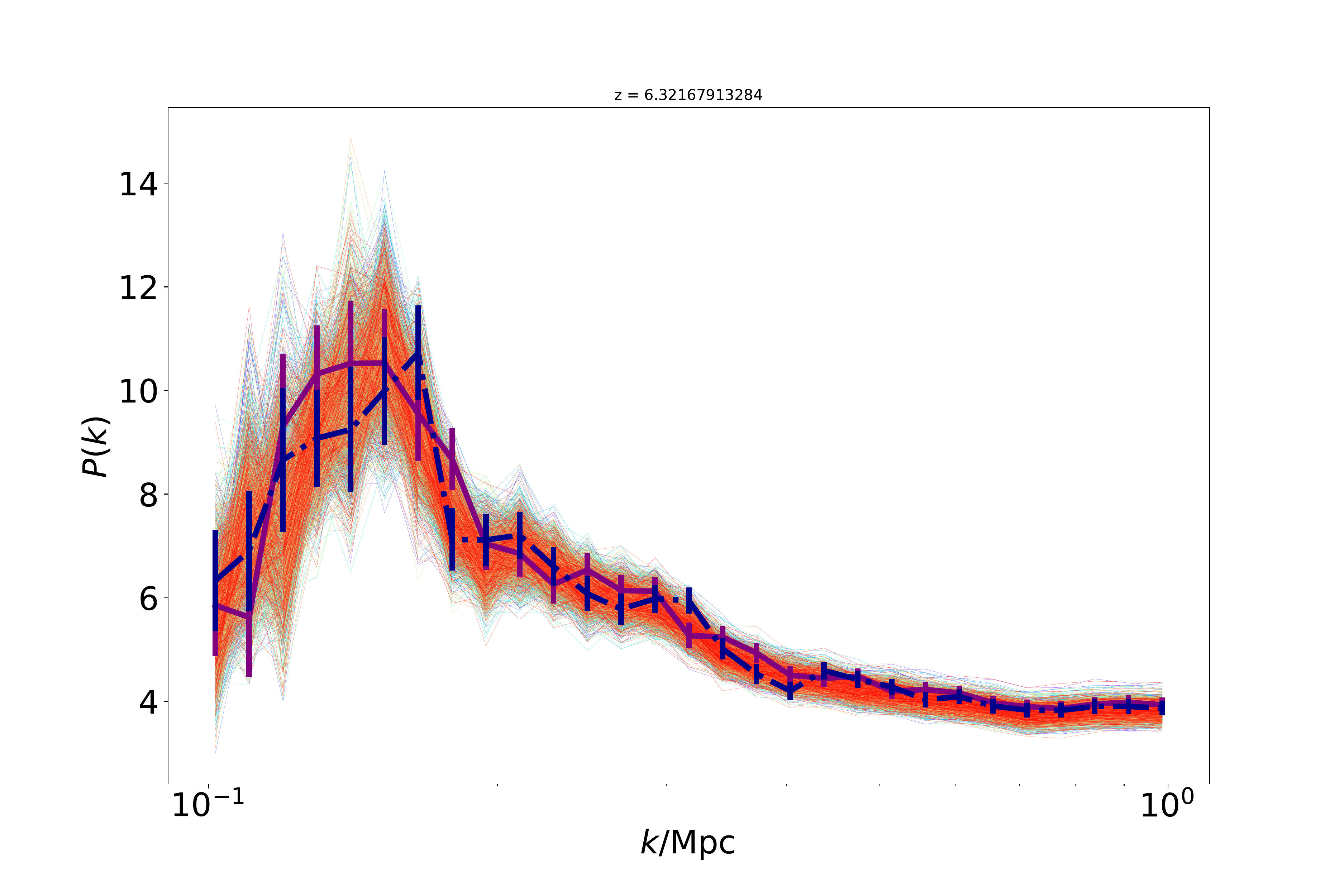}\put(27,115){$z=6.32$}
      \end{overpic}\\
      \begin{overpic}[trim=1.0cm 3cm 3.75cm 2.86cm, clip=true, scale=0.225]{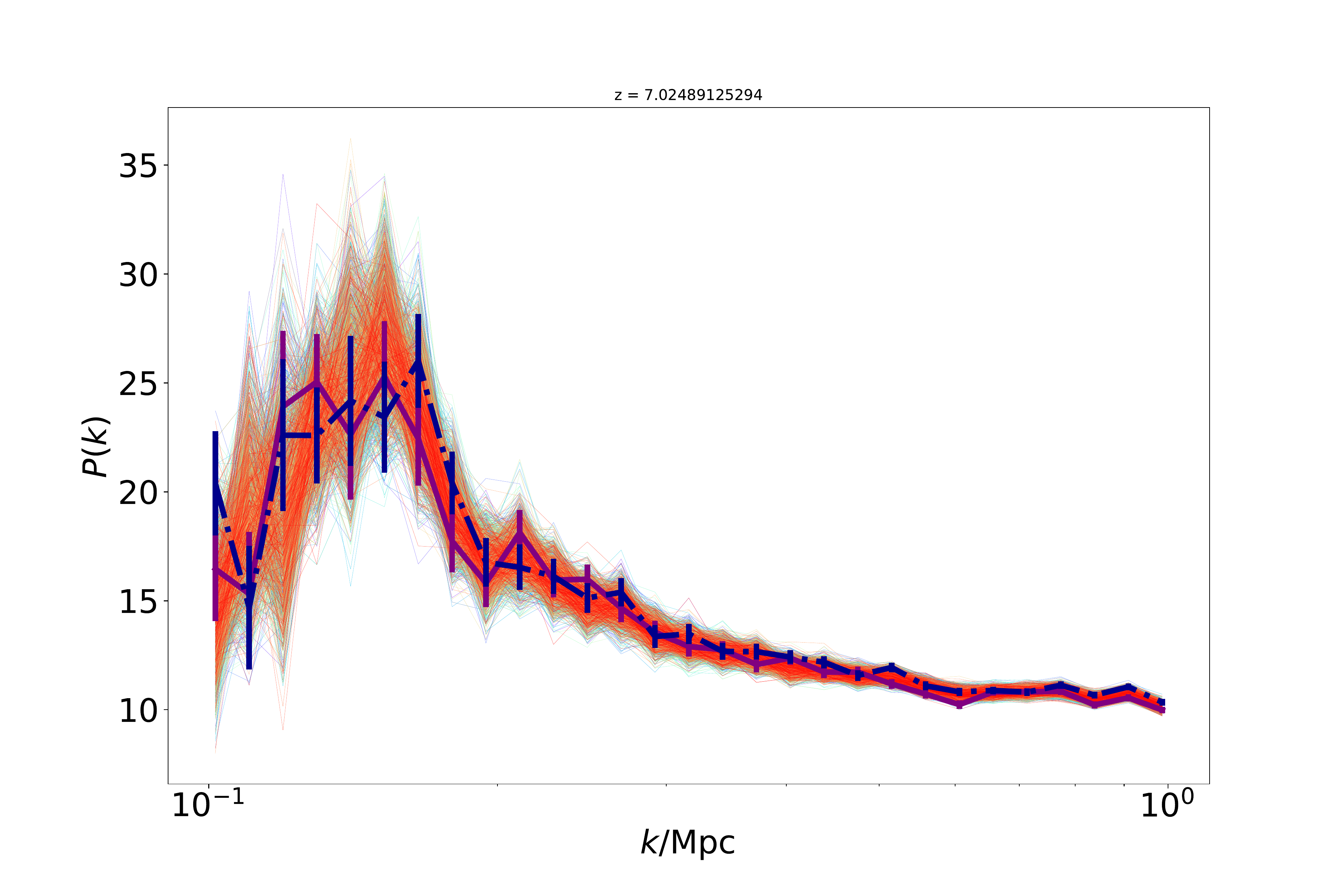}\put(27,115){$ z=7.0$}
      \end{overpic}\\
      \begin{overpic}[trim=1.0cm 3cm 3.75cm 2.85cm, clip=true, scale=0.225]{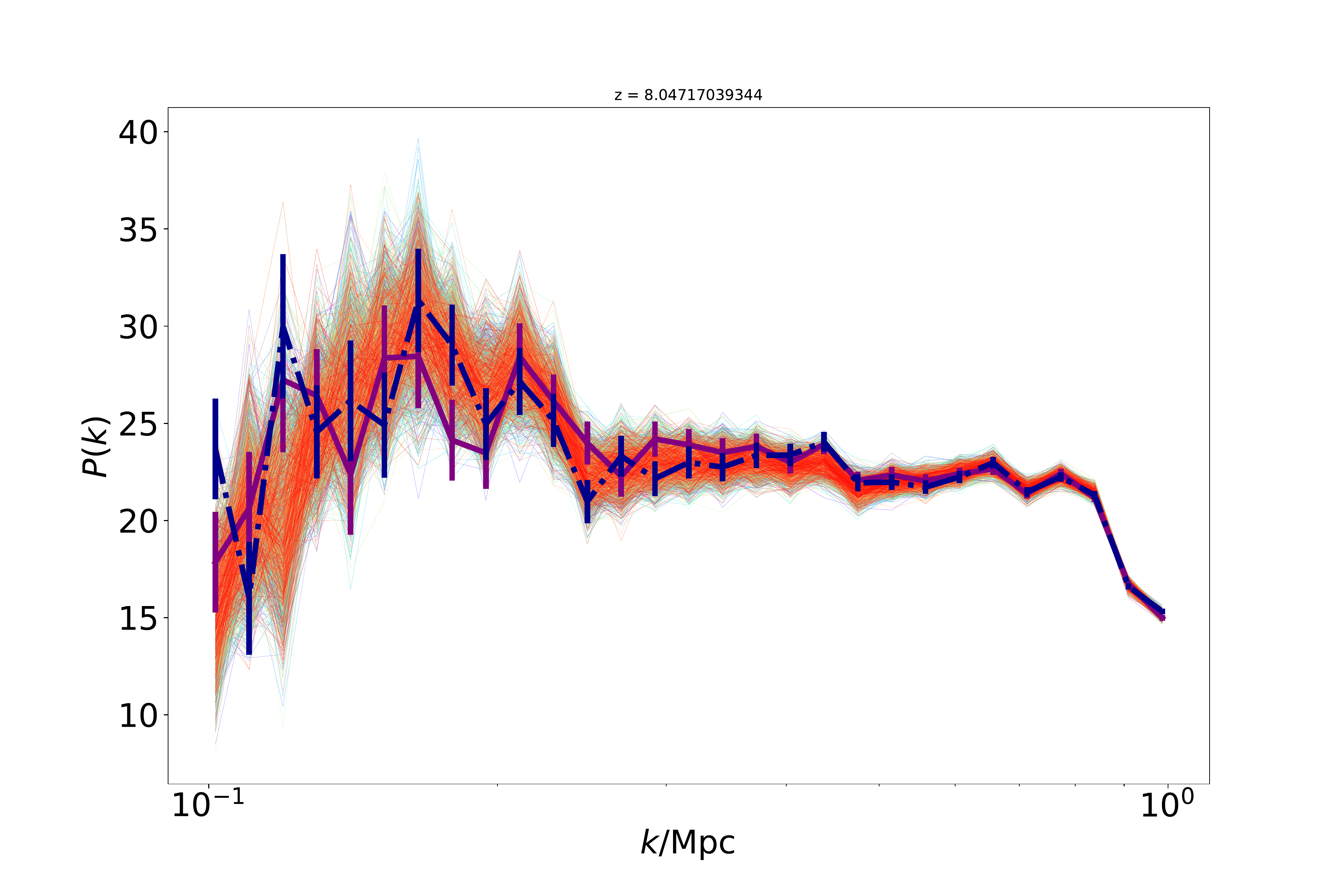}\put(27,115){$z=8.0$}
      \end{overpic}\\
      \begin{overpic}[trim=1.0cm 0.75cm 3.75cm 2.84cm, clip=true, scale=0.225]{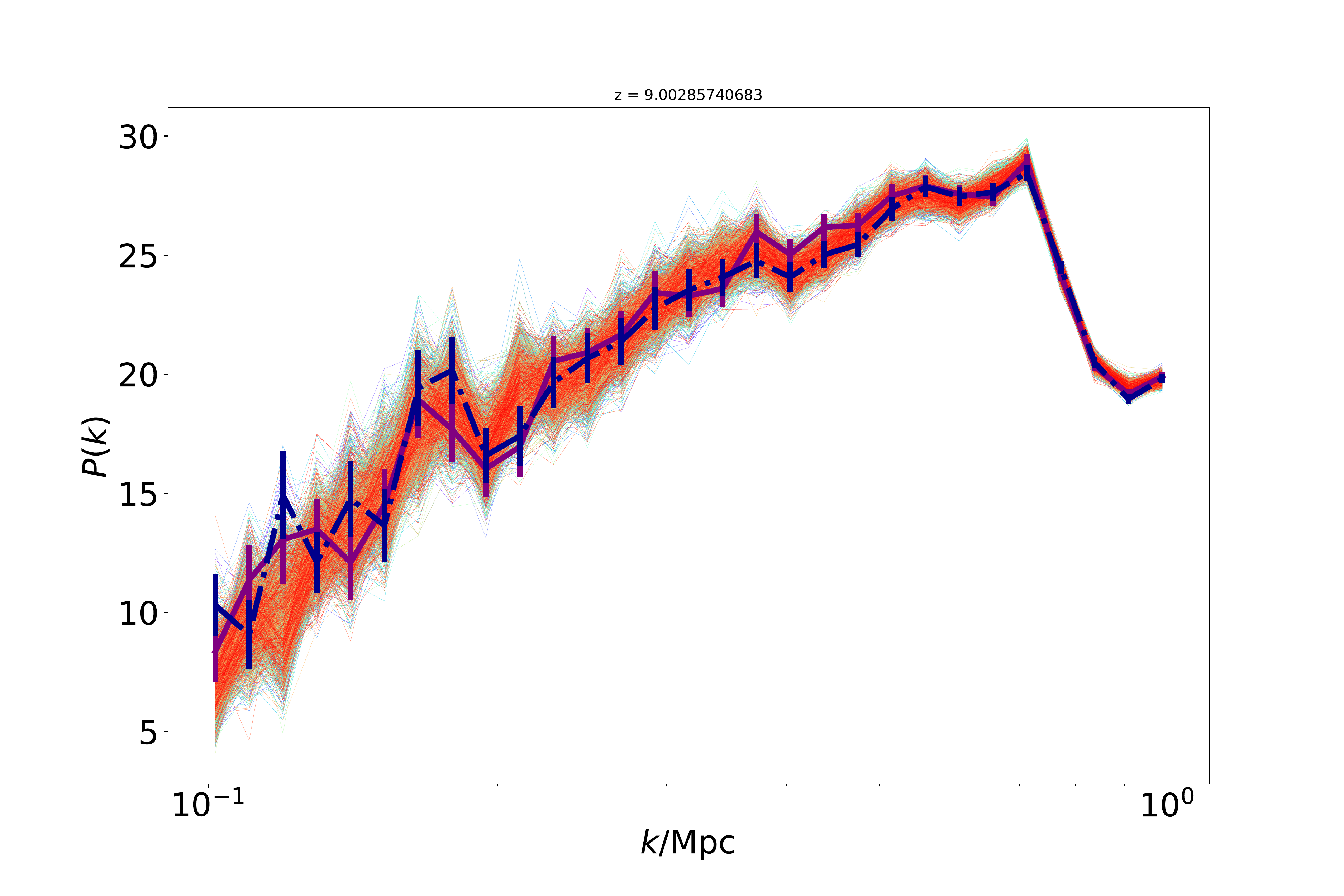}\put(27,130){$z=9.0$}
      \end{overpic}\\
    \end{array}$
  \caption{Here we plot with thin lines all 2000 power spectra used in estimating the error due to sample variance
  for our simulation dimensions.
  The plots from top to bottom correspond to $z = [6.3, 7.0, 8.0, 9.0]$.
  We overplot the two random seeds used in our parameter estimation analysis chosen from about 50 trial runs to minimise (54321) and maximise (6937) the reduced $\chi^2$
  between them and the mean of the distribution of the thin lines shown by the thin lines in the plot.
 }
  \label{fig:ps_random_seeds}
\end{figure}

In Figure \ref{fig:ps_random_seeds} we plot the 2000 power-spectra realisations that we use to calculate our sample-variance errors.
We also overplot the two random seeds used for mock observed data in this study.
As with the bispectrum the extreme seed (purple solid like) is more than $1\sigma$ away from the more standard seed (blue dot-dashed line) for many bins, especially at the later stages of reionization, i.e. $z\le7$.

\bsp
\end{document}